\documentclass[twocolumn]{autart}    

\usepackage{graphicx}          

\usepackage{amssymb}
\usepackage{amsmath}
\usepackage{mathtools}
\usepackage{array}
\usepackage{multirow}
\usepackage[flushleft]{threeparttable}
\usepackage{siunitx}
\usepackage{booktabs}
\usepackage{graphicx}
\usepackage{bm}  
\usepackage{color}   
\usepackage{soul} 
\usepackage{algorithm}
\usepackage{algpseudocode}   
\usepackage{natbib}                      

\newtheorem{definition}{Definition}

\newtheorem{theorem}{Theorem}
\newtheorem{remark}{Remark}
\newtheorem{lemma}{Lemma}

\newtheorem{corollary}{Corollary}

\allowdisplaybreaks[4]

\begin{document}
	
	\begin{frontmatter}
		
		\title{Differentially Private Dual Gradient Tracking \\
				for Distributed Resource Allocation
		} 
        \thanks{The work by W. Huo, X. Chen and L. Shi is supported by the Hong Kong RGC General Research Fund 16211622.
        The work by K. H. Johansson was supported in part by Swedish Research Council Distinguished
Professor Grant 2017-01078, Knut and Alice Wallenberg Foundation Wallenberg Scholar Grant, and Swedish Strategic Research Foundation FUSS SUCCESS Grant. }
		
		
		\author[HKUST]{Wei Huo}\ead{whuoaa@connect.ust.hk},
		\author[HKUST]{Xiaomeng Chen \thanksref{footnoteinfo}}\ead{xchendu@connect.ust.hk},
		\author[NTU]{Lingying Huang}\ead{lingying.huang@ntu.edu.sg},
		\author[KTH]{Karl Henrik Johansson}\ead{kallej@kth.se},
		\author[HKUST]{Ling Shi}\ead{eesling@ust.hk}

		\address[HKUST]{Department of Electronic and Computer Engineering, Hong Kong University of Science and Technology, Hong Kong}
		
		\address[NTU]{School of Electrical and Electronic Engineering, Nanyang Technology University, Singapore 639798}
		
		\address[KTH]{Division of Decision and
			Control Systems, KTH Royal Institute of Technology, Stockholm, Sweden}
		
		\thanks[footnoteinfo]{Corresponding author.}       
		
		\begin{abstract}                          
			This paper investigates privacy issues in distributed resource allocation over directed networks, where each agent holds a private cost function and optimizes its decision subject to a global coupling constraint through local interaction with other agents. 
			Conventional methods for resource allocation over directed networks require all agents to transmit their original data to neighbors, which poses the risk of disclosing sensitive and private information. 
			To address this issue, we propose an algorithm called differentially private dual gradient tracking (DP-DGT) for distributed resource allocation, which obfuscates the exchanged messages using independent Laplacian noise.
			Our algorithm ensures that the agents' decisions converge to a neighborhood of the optimal solution almost surely. 
			Furthermore, without the assumption of bounded gradients, we prove that
			the cumulative differential privacy loss under the proposed algorithm is finite even when the number of iterations goes to infinity. To the best of our knowledge, we are the first to simultaneously achieve these two goals in distributed resource allocation problems over directed networks. 
			Finally, numerical simulations on economic dispatch problems within the IEEE 14-bus system illustrate the effectiveness of our proposed algorithm.
		\end{abstract}
		
		
	\end{frontmatter}
	
\section{Introduction} \label{sec: introduction}	
Resource allocation (RA) is a key issue in fields like smart grids~(\cite{yang2016distributed}) and wireless sensor networks~(\cite{xiao2004simultaneous}), where agents collaboratively optimize their objectives while meeting both local and global constraints. Centralized methods, however, face challenges such as single point failures, high communication demands, and high computational costs~(\cite{wood2013power}), leading to the development of distributed frameworks that rely on agent interactions.
The main challenge in distributed resource allocation (DRA) is managing global resource constraints that link all agents' decisions. \cite{zhang2012convergence} introduced a leader-follower consensus algorithm for quadratic objectives. To address the limitation of quadratic cost functions, \cite{yi2016initialization} proposed a primal-dual algorithm. These works focused on undirected networks with doubly stochastic information mixing matrices. However, bidirectional information flows may incur unnecessary communication costs or may not exist due to sensor power heterogeneity.
To tackle this, \cite{yang2016distributed} introduced a distributed algorithm for unbalanced directed networks using the gradient push-sum method, though it required agents to calculate the eigenvector of the weight matrix asymptotically. Later, \cite{zhang2020distributed} explored the dual relationship between DRA and distributed optimization (DO), employing the push-pull technique for determining explicit convergence rates.
	
The existing literature assumes secure transmission of raw data, but the distributed nature of cyber-physical systems raises privacy concerns. Messages between agents are vulnerable to interception, risking theft or inference of sensitive data. For example, in smart grid economic dispatch problems, transmitted messages could expose private user patterns or financial information. Thus, privacy-preserving algorithms for distributed resource allocation are essential.
While encryption prevents eavesdropping~(\cite{freris2016distributed}), it may not be feasible for large-scale distributed systems with limited battery power. \cite{beaude2020privacy} used secure multi-party computation for privacy protection in DRA, but this does not defend against external attackers who intercept all messages. It is also inefficient due to the complexity of computation and communication.
\cite{lu2023privacy} applied conditional noise to protect DRA privacy, but their analysis was limited to quadratic cost functions and only defined privacy as preventing the inference of the exact cost function, which is too narrow and doesn't cover all scenarios. In contrast, differential privacy (DP) has gained attention for its strong mathematical guarantees and robustness under post-processing~(\cite{dwork2006differential}).
	
Recent studies have developed differentially private algorithms for unconstrained distributed optimization (DO) over directed networks. \cite{chen2023differentially} used state-decomposition and constant noise for privacy-preserving DO, but this only preserves privacy per iteration, leading to cumulative privacy loss over time. Their convergence and privacy analysis also relied on gradient boundedness, which is impractical for applications like distributed economic dispatch with quadratic cost functions. \cite{wang2023tailoring} relaxed the gradient boundedness assumption, ensuring $\epsilon$-DP over infinite iterations, but required adjacency gradients to be identical near the optimal point. \cite{huang2024differential} showed that gradient tracking-based algorithms cannot achieve $\epsilon$-DP under Laplacian noise with non-summable step sizes. These studies assume strong convexity or convexity of the cost functions. For non-convex objectives, privacy-preserving algorithms have been proposed by \cite{wang2022decentralized} and \cite{wang2023decentralized}, but they rely on bounded gradient assumptions and apply only to undirected graphs.
For DRA with global coupling constraints, \cite{han2016differentially} achieved DP by adding noise to public signals, but this requires an additional entity for information collection and broadcasting. \cite{ding2021differentiallyra} preserved the privacy of cost functions for DRA, though their work is limited to undirected networks. \cite{wang2024robust} used primal-dual algorithms for finite DP in DRA, but their approach also applies only to undirected graphs and requires additional variables and noise for directed graphs. For directed networks, \cite{lu2023privacy} employed conditional noise for privacy protection in distributed economic dispatch problems. However, their privacy analysis is limited to quadratic cost functions, and their definition of privacy lacks flexibility and generality. To date, no work addresses differentially private DRA over directed networks.
	
Motivated by the observations above, our work focuses on providing DP for DRA over directed graphs. We specifically target $\delta$-adjacent DRA problems (\textbf{Definition~\ref{defn: adjacency}}), which relaxes the bounded gradient assumption.
To address global coupling constraints, we consider the dual formulation of DRA. While the dual relationship between DO and DRA, along with existing privacy-preserving DO algorithms for directed networks, is well-established, it is important to note that the dual objective function in DRA is not always strongly convex. As a result, the analysis methods in \cite{pu2020robust} and \cite{huang2024differential} are not directly applicable.
We thus analyze the convergence of gradient-tracking with noisy shared information for \emph{non-convex} objectives, significantly extending the analysis in previous works~(\cite{chen2023differentially, pu2020robust, wang2023tailoring, huang2024differential}). We derive conditions on the step size and noise to ensure both convergence and $\epsilon$-DP over infinite iterations. A comparison of key related works is presented in Table~\ref{tab: compare}.
	\begin{table*}[t] 
		\caption{A Comparison of Some Related Decentralized Algorithms.
		}
		\label{tab: compare}
        \tiny
		\centering
		\begin{tabular}{|m{0.14\textwidth}<{\centering}|m{0.11\textwidth}<{\centering}|m{0.07\textwidth}<{\centering}|m{0.32\textwidth}<{\centering}|m{0.22\textwidth}<{\centering}|}
			\hline
			 & Problem & Topology & Gradient Assumption & DP Consideration  \\
			 \hline
			\cite{zhang2020distributed} & DRA & Directed & No assumption & $\times$ \\
			\hline
			\cite{ding2021differentiallyra} & DRA & Undirected & Same adjacent gradients in the horizontal position & $\epsilon$-DP over infinite iterations \\
            \hline
			\cite{wang2022decentralized} & DO (non-convex) & Undirected & Uniformly bounded gradient & Information-theoretic privacy at each  iteration\\
			\hline
		\cite{wang2023decentralized} & DO (non-convex) & Undirected & Uniformly bounded gradient & $(\epsilon,\delta)$-DP at each  iteration \\
			\hline
			\cite{chen2023differentially} & DO (strongly convex) & Directed & Uniformly bounded gradient & $\epsilon$-DP over finite iterations \\
			\hline
			\cite{wang2023tailoring} & DO (convex) & Directed & Same adjacent gradients near the optimal point & $\epsilon$-DP over infinite iterations \\
			\hline
		    \cite{huang2024differential} & DO (strongly convex) & Directed & Bounded distance between adjacent gradients & $\epsilon$-DP over infinite iterations \\
		    \hline
			Our work & DRA & Directed & Bounded distance between adjacent gradients & $\epsilon$-DP over infinite iterations \\
			\hline
		\end{tabular}
	\end{table*}
	In summary, our main contributions are as follows:
	\begin{itemize}
		\item[1)] We propose a differentially private dual gradient tracking algorithm, abbreviated
		as DP-DGT (\textbf{Algorithm~\ref{algo: one}}), to address privacy issues in DRA over directed networks. Our algorithm masks the transmitted messages in networks with Laplacian noise and does not rely on any extra central authority.
		\item[2)] With the derived sufficient conditions, we prove that the DP-DGT algorithm converges to a neighborhood of the optimal solution (\textbf{Theorem~\ref{thm: primal_con}}) by showing the convergence of the dual variable even for non-convex objectives (\textbf{Theorem~\ref{thm: convergence}}). This theoretical analysis nontrivially extends existing works on gradient tracking with information-sharing noise for convex or strongly convex objectives~(\cite{chen2023differentially, wang2023tailoring, pu2020robust, huang2024differential}).
		\item[3)] We specify the mathematical expression of privacy loss $\epsilon$ under the DP-DGT (\textbf{Theorem~\ref{thm: DP}}) and demonstrate that the DP-DGT algorithm preserves DP for each individual agent's cost function even over infinite iterations (\textbf{Corollary~\ref{cor: finite}}).
		To our best knowledge, previous studies have only reported differential privacy results for DRA in undirected networks~(\cite{ding2021differentiallyra}). Moreover, our analysis relaxes the gradient assumption used in~\cite{wang2023tailoring, chen2023differentially}.
	\end{itemize}
	
The remainder of this paper is structured as follows: Section~\ref{sec: formulation} introduces the preliminaries and problem formulation for privacy-preserving DRA over directed networks. Section~\ref{sec: algo} presents a differentially private distributed dual gradient tracking algorithm with robust push-pull. Section~\ref{sec: convergence} details the convergence analysis, followed by a rigorous proof of $\epsilon$-DP over infinite iterations in Section~\ref{sec: DP}. Section~\ref{sec: sim} provides numerical simulations illustrating the results, and Section~\ref{sec: conclusions} discusses conclusions and future research directions.
	
	\emph{Notations}:
	Let $\mathbb{R}^{p}$ and $\mathbb{R}^{p \times q}$ represent the set of $p$-dimensional vectors and $p \times q$-dimensional matrices, respectively.
	The notation $\mathbf{1}_{p} \in \mathbb{R}^{p}$ denotes a vector with all elements equal to one, and $I_{p} \in \mathbb{R}^{p \times p}$ represents a $p \times p$-dimensional identity matrix.
	We use $\|\cdot \|_{2}$ or $\| \cdot \|$ to denote the $\ell_{2}$-norm of vectors and the induced $2$-norm for matrices.
	We use $\mathbb{P}(\mathcal{A})$ to represent the probability of an event $\mathcal{A}$, and $\mathbb{E}[x]$ to be the expected value of a random variable $x$.
	The notation $\text{Lap}(\theta)$ denotes the Laplace distribution with probability density function $f_{L}(x|\theta) = \frac{1}{2\theta}e^{\frac{-|x|}{\theta}}$, where $\theta > 0$.
	If $x \sim \text{Lap}(\theta)$, we have $\mathbb{E}[x^{2}] = 2 \theta^{2}$ and $\mathbb{E}[x] = 0$.
	
	\emph{Graph Theory}:
	A directed graph is denoted as $\mathcal{G} = (\mathcal{N}, \mathcal{E})$, where $\mathcal{N} = \{1, 2, \dots, N\}$ is the set of nodes and $\mathcal{E} \subseteq \mathcal{N} \times \mathcal{N}$ is the edge set consisting of ordered pairs of nodes. 
	Given a nonnegative matrix $\mathbf{A} = [a_{ij}] \in \mathbb{R}^{N \times N}$, the directed graph induced by $\mathbf{A}$ is referred to as $\mathcal{G}_{\mathbf{A}} = (\mathcal{N}, \mathcal{E}_{\mathbf{A}})$, where the directed edge $(i,j)$ from node $j$ to node $i$ exists, i.e., $(i,j) \in \mathcal{E}_{\mathbf{A}}$ if and only if $a_{ij} >0$.
	For a node $i \in \mathcal{N}$, its in-neighbor set $\mathcal{N}_{\mathbf{A},i}^{\text{in}}$ is defined as the collection of all individual nodes from which $i$ can actively and reliably pull data in graph $\mathcal{G}_{\mathbf{A}}$. Similarly, its out-neighbor set $\mathcal{N}_{\mathbf{A},i}^{\text{out}}$ is defined as the collection of all individual agents that can passively and reliably receive data from node $i$.
	
	\section{Preliminaries and Problem Statement} \label{sec: formulation}
This section provides the preliminaries and problem formulations. We first introduce the RA problem over networks and its dual counterpart. Next, we outline the class of algorithms considered and the messages exchanged. We then discuss potential privacy concerns in traditional algorithms and introduce concepts related to DP. Finally, we formulate the problems addressed in this work.
	
	\subsection{Resource Allocations over Networks}
	We consider a network of $N$ agents that interact on a directed graph $\mathcal{G}$ to collaboratively address a RA problem. Each agent $i$ possesses a local private cost function $F_{i}: \mathbb{R}^{m} \to \mathbb{R}$. Their goal is to solve the following resource allocation problem using a distributed algorithm over $\mathcal{G}$:
	\begin{equation} \label{eq: problem}
		\begin{aligned}
			\min_{\mathbf{w} \in \mathbb{R}^{N \times m}} & \quad F(\mathbf{w}) = \sum_{i=1}^{N}F_{i}(w_{i}), \\
			s.t. & \quad \sum_{i=1}^{N} w_{i} = \sum_{i=1}^{N} d_{i}, \ w_{i} \in \mathcal{W}_{i}, \ \forall i \in \mathcal{N},
		\end{aligned}
	\end{equation}
	where $w_{i} \in \mathbb{R}^{m}$ represents the local decision of agent $i$, indicating the resource allocated to the agent, $\mathcal{W}_{i} \subseteq \mathbb{R}^{m}$ refers to the local closed and convex constraint set, $\mathbf{w} = [w_{1}, \dots, w_{N}]^{T} \in \mathbb{R}^{N \times m}$, and $d_{i}$ denotes the local private resource demand of agent $i$. 
	Let $d = \sum_{i=1}^{N}d_{i}$, and thus $\sum_{i=1}^{N}w_{i} = d$ represents the overall balance between supply and demand, indicating the coupling among agents.
	
	Throughout the paper, we make the following assumptions:
	\begin{assum} \label{assum: convexity_slater}
		(Strong convexity and Slater's condition):
		\vspace{-0.3cm}
		\begin{itemize}
			\item[1)] The local cost function $F_{i}$ is $\mu$-strongly convex for all $i \in \mathcal{N}$, i.e., for any $w, w^{\prime} \in \mathbb{R}^{m}$, $\| \nabla F_{i}(w) - \nabla F_{i}(w^{\prime}) \| \geq \mu \| w - w^{\prime} \|$.
			\item[2)] There exists at least one point in the relative interior $\mathcal{W}$ that can satisfy the power balance constraint $\sum_{i=1}^{N} w_{i} = \sum_{i=1}^{N} d_{i}$, where $\mathcal{W} = \mathcal{W}_{1} \times \cdots \times \mathcal{W}_{N}$.
		\end{itemize}
	\end{assum}
	
	Assumption~\ref{assum: convexity_slater} ensures strong duality between problem~\eqref{eq: problem} and its dual counterpart, enabling us to deal with the coupling constraint based on its dual problem.
	
	\subsection{Dual Problem}
	To handle the global constraint, we begin by formulating the dual problem of~\eqref{eq: problem}.
	The Lagrange function of~\eqref{eq: problem} is given by
	\begin{equation} \label{eq: lang}
		\mathcal{L}(\mathbf{w}, x) =  \sum_{i=1}^{N} F_{i}(w_{i}) + x^{T}\left( \sum_{i=1}^{N}w_{i} - \sum_{i=1}^{N} d_{i} \right),
	\end{equation}
	where $x \in \mathbb{R}^{m}$ represents the dual variable. 
	Thus, the dual problem of~\eqref{eq: problem} can be expressed by 
	\begin{equation} \label{eq: dual}
		\max_{x \in \mathbb{R}^{m}} \ \inf_{\mathbf{w} \in \mathcal{W}} \mathcal{L}(\mathbf{w}, x).
	\end{equation} 
	The objective function in~\eqref{eq: dual} can be written as
	\begin{align*}
		&\inf_{\mathbf{w} \in \mathcal{W}} \mathcal{L}(\mathbf{w}, x) \\
		=& \sum_{i=1}^{N} \inf_{w_{i} \in \mathcal{W}_{i}} \left( F_{i}(w_{i}) + x^{T} w_{i} \right) - x^{T} \sum_{i=1}^{N}d_{i} \\
		=& \sum_{i=1}^{N} -F_{i}^{*}(-x) - x^{T} \sum_{i=1}^{N}d_{i},
	\end{align*}
	where
	\begin{equation} \label{eq: sup}
		F_{i}^{*}(x) = \sup_{w_{i} \in \mathcal{W}_{i}} \left( x^{T}w_{i} - F_{i}(w_{i}) \right)
	\end{equation} 
	corresponds to the convex conjugate function for the pair $(F_{i}, \mathcal{W}_{i})$~(\cite{bertsekas1997nonlinear}).
	Consequently, the dual problem~\eqref{eq: dual} can be formulated as the subsequent DO problem
	\begin{equation} \label{eq: dual_form}
		\min_{\mathbf{x} \in \mathbb{R}^{m}} \ f(x)= \sum_{i=1}^{N}f_{i}(x), \ \text{where} \ f_{i}(x) \triangleq  F_{i}^{*}(-x) + x^{T} d_{i}.
	\end{equation}
	According to the Fenchel duality between strong convexity and the Lipschitz continuous gradient~(\cite{zhou2018fenchel, boyd2004convex}), the strong convexity of $F_{i}$ indicates the differentiability of $F_{i}^{*}$ with Lipschitz continuous gradients, and the supremum in~\eqref{eq: sup} can be attained. 
	Danskin's theorem states that $\nabla F_{i}^{*}(x) = \mathop{\arg\max}\limits_{w \in \mathcal{W}_{i}}\{ x^{T}w - F_{i}(w) \}$~\cite{bertsekas1997nonlinear}, providing the gradient of $F_{i}^{*}$. 
	Therefore, we have
	\begin{equation}
		\begin{aligned}
			\nabla f_{i}(x) &= -\nabla F_{i}^{*}(-x) + d_{i} \\
			&= -\mathop{\arg\min}\limits_{w \in \mathcal{W}_{i}} \left\{ F_{i}(w)+x^{T}w \right\} + d_{i}.
		\end{aligned}
	\end{equation}
	We find that the dual gradient $\nabla f_{i}(x)$ captures the local deviation or mismatch between resource supply and demand, i.e., $w_{i} - d_{i}$, in some sense.
	
	If Assumption~\ref{assum: convexity_slater} holds, we observe the strong duality between the dual DO problem~\eqref{eq: dual} and the primal DRA problem~\eqref{eq: problem}.
	This equivalence is captured through $F^{*} = -f^{*}$ and the optimal solution $x^{*}$ of~\eqref{eq: dual} satisfies $F_{i}^{*}(w_{i}^{*}) + F_{i}^{*}(-x^{*}) = - x^{*T}w_{i}^{*}$. 
	As a result, if the proposed algorithm can drive the dual variable in~\eqref{eq: dual_form} to the optimal one, it equivalently steers $\mathbf{w}$ to the optimal solution $\mathbf{w}^{*}$ in~\eqref{eq: problem}.
	Hence, our algorithmic focus can be directed towards solving~\eqref{eq: dual_form}, the standard DO problem.
	
	\subsection{Communication Networks and Information Flows}
	
	Let us introduce the class of DRA algorithms we are considering in this paper. Gradient-tracking with push-pull~(\cite{pu2020push}) is one of the DO algorithms that enable agents to solve optimization problems in directed networks, particularly for those unbalanced networks lacking doubly stochastic weight matrices. 
	Solving the dual counterpart~\eqref{eq: dual_form} using this algorithm allows agents to obtain the optimal solution for the DRA problem~\eqref{eq: problem} over $\mathcal{G}$.
	Specifically, agent $i$ maintains a local estimate of the dual variable $x$ and a local estimate of the global constraint deviation at iteration $k$, denoted as $\tilde{w}_{i,k}$ and $z_{i,k}$, respectively.
	These two local variables are shared using two different communication networks, $\mathcal{G}_{\mathbf{R}}$ and $\mathcal{G}_{\mathbf{C}^{T}}$, respectively.
	These two networks are induced by matrices $\mathbf{R} = [R_{ij}] \in \mathbb{R}^{N \times N}$ and $\mathbf{C} = [C_{ij}] \in \mathbb{R}^{N \times N}$, respectively, where $R_{ij}>0$ for any $j \in \mathcal{N}_{\mathbf{R}, i}^{\text{in}}$ and $C_{ij}>0$ for any $i \in \mathcal{N}_{\mathbf{C}, j}^{\text{out}}$.
	We call this special class of algorithms \emph{DRA with dual gradient tracking}, denoted as $\mathcal{A}$. A representative form of $\mathcal{A}$ that we consider in this paper is as follows~(\cite{zhang2020distributed}):
	\begin{subequations} \label{eq: or_alg1}
		\begin{align}
			\label{eq: wbar_or} \tilde{w}_{i,k+1} =&  \sum_{j=1}^{N}R_{ij} \tilde{w}_{j,k} + \beta_{k} z_{i,k}, \\
			\label{eq: w_or} w_{i,k+1} =& \mathop{\arg\min}\limits_{w \in \mathcal{W}_{i}} \left\{ F_{i}(w) - \tilde{w}_{i,k+1}^{T}w \right\}, \\
			\label{eq: z_or}	 z_{i,k+1} =&  \sum_{j=1}^{N}C_{ij} z_{j,k} - \iota (w_{i,k+1} - w_{i,k}),
		\end{align}
	\end{subequations}
	where $\beta_{k}>0$ and $\iota > 0$ are the step sizes. 
	\begin{figure}[t]  
		\centering
		\includegraphics[width=0.8\linewidth]{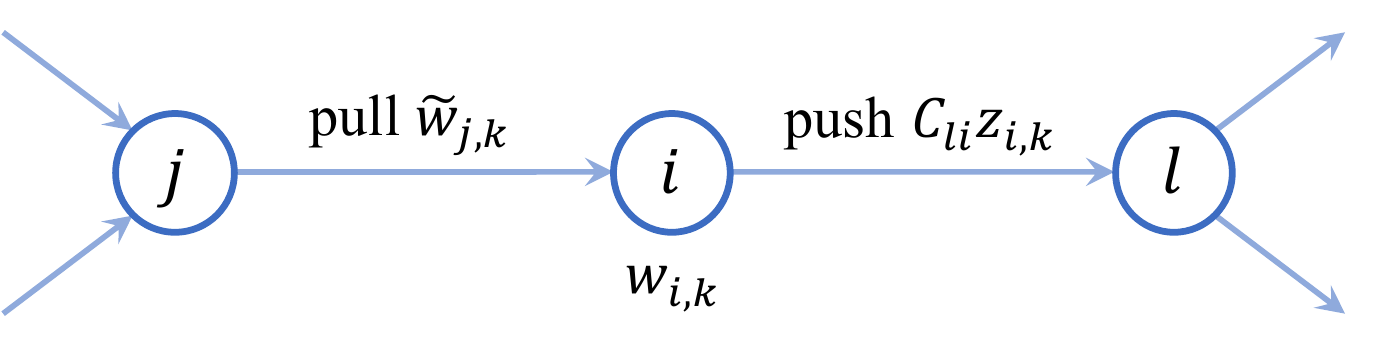}
		\caption{Information flows of agent $i$ under algorithm $\mathcal{A}$.} 
		\label{fig: push_pull}
	\end{figure}
	Fig.~\ref{fig: push_pull} illustrates the information flows under algorithm $\mathcal{A}$ over the communication network. At each iteration, agent $i$ pushes $C_{li}z_{i,k}$ to each out-neighboring agent $l \in \mathcal{N}_{\mathbf{C},i}^{\text{out}}$, pulls the dual variable estimate $\tilde{w}_{j,k}$ from its in-neighboring agent $j \in \mathcal{N}_{\mathbf{R},i}^{\text{in}}$, and updates its privately-owned primal optimization variable $w_{i,k}$.
	
	We impose the following assumptions on the communication graphs:
	\begin{assum} \label{assum: spanning_tree}
		The graphs $\mathcal{G}_{\mathbf{R}}$ and $\mathcal{G}_{\mathbf{C}^{T}}$ each contains at least one spanning tree. Moreover, there exists at least one node that is a root of a spanning tree for both $\mathcal{G}_{\mathbf{R}}$ and $\mathcal{G}_{\mathbf{C}^{T}}$.
	\end{assum}
	\vspace{-0.2cm}
	\begin{assum} \label{assum: rc_stochastic}
		The matrix $\mathbf{R}$ is row-stochastic and $\mathbf{C}$ is column-stochastic, i.e., $\mathbf{R}\mathbf{1} = \mathbf{1}$ and $\mathbf{1}^{T}\mathbf{C} = \mathbf{1}^{T}$.
	\end{assum}
	
	Assumption~\ref{assum: spanning_tree} is less restrictive than previous works such as~\cite{tsianos2012push} and \cite{xi2017dextra}, as it does not necessitate a strongly connected directed graph. This allows more flexibility in graph design. 
	However, directly transmitting $\tilde{w}_{j,k}$ and $C_{li}z_{i,k}$ in the network will pose privacy concerns, which we will discuss in the next subsection.

	\subsection{Differential Privacy}
	
	In deterministic optimization problems, given a specific initialization and topology, the generated data and decisions are uniquely determined by the cost function of each agent. Therefore, in insecure networks, agents should protect the privacy of their cost functions against eavesdroppers while calculating the optimal solution to~\eqref{eq: problem} in a distributed manner.  
	In this paper, we consider the following commonly used eavesdropping attack model~(\cite{wang2023tailoring, chen2023differentially}):
	\begin{definition}
		An eavesdropping attack is an adversary that is able to listen to all communication messages in the network.
	\end{definition}
	Note that our definition of eavesdropping models a powerful attack as the adversary potentially can intercept every message in the network. 
	For example, in the communication network shown in Fig.~\ref{fig: push_pull}, an adversary under eavesdropping attack can obtain $\{\tilde{w}_{i,k}, C_{li}z_{i,k}| \forall i \in \mathcal{N}, k \geq 0 \}$.
	With this observation, the attacker is able to learn $w_{i,k}$ based on publicly known $\mathbf{R}$, $\mathbf{C}$, and step sizes.  
	If the local constraint set $\mathcal{W}_{i}$ is equal to $\mathbb{R}^{m}$ and $F_{i}$ is differentiable, the step \eqref{eq: w_or} can be rewritten as $w_{i,k+1} = \nabla^{-1} F_{i}(\tilde{w}_{i,k+1})$, where $\nabla^{-1} F_{i}$ represents the inverse function of $\nabla F_{i}$ such that $\nabla F_{i}(w_{i,k+1}) = \tilde{w}_{i,k+1}$. 
	As a result, the cost function $F_{i}$ could be deduced from~\eqref{eq: w_or}, which causes privacy leakage. 
	
	The potential privacy leakage in algorithms in $\mathcal{A}$ motivates us to design novel privacy-preserving algorithms to protect the privacy of agents' cost functions.
	To measure privacy, we introduce concepts associated with DP~(\cite{dwork2006differential}).
	
	First, we formulate the closeness of DRA problems.  
	We denote the DRA problem shown in~\eqref{eq: problem} as $\mathcal{P}$ and represent it by four parameters $(\mathcal{W}, \mathcal{S}, F, \mathcal{G} )$, where $\mathcal{S} \subseteq \{\mathbb{R}^{m} \to  \mathbb{R} \} $ is a set of real-valued cost functions, and $F = \sum_{i=1}^{N}F_{i}$ with $F_{i} \in \mathcal{S}$ for each $i \in \mathcal{N}$.
	Specifically, we define $\delta$-adjacency of two DRA problems by measuring the distance between gradients of the individual's local cost function.

	\begin{definition} \label{defn: adjacency} ($\delta$-adjacency)
		Two distributed resource allocation problems $\mathcal{P}$ and $\mathcal{P}^{\prime}$ are $\delta$-adjacent if the following conditions hold:
		\vspace{-0.2cm}
		\begin{itemize}
			\item[1)] they share identical domains for resource allocation and communication graphs, i.e., 
			$\mathcal{W} = \mathcal{W}^{\prime}$ and $\mathcal{G} = \mathcal{G}^{\prime}$;
			\item[2)] there exists an $i_{0} \in \mathcal{N}$ such that $F_{i_{0}} \neq  F^{\prime}_{i_{0}}$, and for all $i \neq i_{0}$, $F_{i} = F^{\prime}_{i}$;
			\item[3)] the distance of gradients of $F_{i_{0}}$ and $F^{\prime}_{i_{0}}$ are bounded by $\delta$ on $\mathcal{W}_{i_{0}}$, i.e., $\sup\limits_{w \in \mathcal{W}} \| \nabla F_{i_{0}}(w) - \nabla F_{i_{0}}^{\prime}(w) \| \leq \delta$ for any $w \in \mathcal{W}_{i_{0}}$.
		\end{itemize} 
	\end{definition}
	\begin{remark}
		According to Definition~\ref{defn: adjacency}, two DRA problems are adjacent when the cost function of a single agent changes, while all other conditions remain unchanged. As shown in Table~\ref{tab: compare}, various works define $\delta$-adjacency differently for DP analysis. Many earlier studies relied on the bounded gradient assumption, but this fails for quadratic cost functions commonly used in resource allocation problems~(\cite{yang2016distributed, kar2012distributed, lu2023privacy}). Our assumption relaxes this requirement, allowing for more general cost functions. Additionally, unlike \cite{wang2023tailoring}, we do not require adjacent gradients to be identical, nor do we require cost functions to be the same around the optimal value, as in \cite{wang2023decentralized, wang2024robust}.
	\end{remark}

\begin{remark} 
In this paper, we focus on adjacent functions with different objective functions but identical coupling constraints, aiming to protect the private information in each agent's local objective function, denoted as $f_i$. This approach is common in privacy-preserving distributed resource allocation problems, such as those in \cite{ding2021differentially} and \cite{hu2023achieving}. Works that address private coupling constraints, such as \cite{munoz2021private}, are beyond the scope of our study.
\end{remark}
	Traditional algorithms in $\mathcal{A}$ do not have any privacy protection in general~(\cite{zhang2020distributed}). To preserve the privacy of agents, we should add some random perturbations or uncertainties to confuse the eavesdropper. 
	We denote the class of algorithms $\mathcal{A}$ with random perturbations as $\mathcal{R}$.
	Given a DRA problem $\mathcal{P}$, an iterative randomized algorithm can be considered as a mapping as $\mathcal{R}_{\mathcal{P}}(v_{0}): v_{0} \to \mathcal{O}$, where $v_{0}$ is the set of initial states of the algorithm $\mathcal{R}$ and $\mathcal{O}$ is the observation sequence of all shared messages.
	
	Let us now define privacy for such a randomized algorithm following the classical $\epsilon$-DP notion introduced by \cite{dwork2006differential}.
	
	\begin{definition} \label{defn: DP} ($\epsilon$-DP)
		For a given $\epsilon > 0$, a randomized iterative algorithm $\mathcal{R}$ solving~\eqref{eq: problem} is $\epsilon$-DP if for any two $\delta$-adjacent resource allocation problems $\mathcal{P}$ and $\mathcal{P}^{\prime}$, any set of observation sequences $\mathcal{O} \subseteq \emph{Range}(\mathcal{R})$\footnote{$\text{Range}(\mathcal{R})$ denotes the set of all possible observation sequences under the algorithm $\mathcal{R}$.} and any initial state $v_{0}$, it holds that
		\begin{equation}
			\mathbb{P}[\mathcal{R}_{\mathcal{P}}(v_{0}) \in \mathcal{O} ] \leq e^{\epsilon} \mathbb{P}[\mathcal{R}_{\mathcal{P}^{\prime}}(v_{0})) \in \mathcal{O} ],
		\end{equation}
		where the probability is over the randomness introduced in each iteration of the algorithm.
	\end{definition}
Definition~\ref{defn: DP} defines $\epsilon$-differential privacy (DP) for a randomized algorithm $\mathcal{R}(\cdot)$, ensuring minimal differentiation in output probabilities for two $\delta$-adjacent RA problems. Here, $\epsilon$ represents the privacy budget or loss, with smaller values making it harder for an eavesdropper to distinguish between two sets of cost functions based on observed data. However, in many existing works~(\cite{chen2023differentially, wang2022quantization, chen2022fundamental}), a finite cumulative privacy budget is only achieved over a limited number of iterations. As iterations increase, even if the solution nears optimal, privacy may eventually be compromised. Thus, we carefully design the step size and noise parameters to preserve privacy.

	\subsection{Problem Formulation}

This work first designs a novel distributed algorithm $\mathcal{R}$ that preserves privacy for RA problems on directed graphs. We then analyze the conditions on step sizes and perturbations that ensure both convergence and $\epsilon$-DP over infinite iterations.
	
	\section{Algorithm Development} \label{sec: algo}
DP is typically maintained by introducing noise into transmitted data. However, information-sharing noise corrupts exchanged messages, causing agents to receive distorted estimates of dual variables and constraint deviations, which reduces accuracy. This creates a fundamental trade-off between privacy and optimization accuracy. 
	To understand the impact of noise, we start by analyzing the update of $\tilde{w}_{i,k}$ and $z_{i,k}$ under the traditional algorithm~\eqref{eq: or_alg1} in the presence of information-sharing noise.
	
	Defining $\tilde{\mathbf{w}}_{k} = [\tilde{w}_{1,k}, \dots, \tilde{w}_{N,k}]^{T} \in \mathbb{R}^{N \times m}$ and $\mathbf{z}_{k} = [z_{1,k}, \dots, z_{N,k}]^{T} \in \mathbb{R}^{N \times m}$, we can rewrite~\eqref{eq: wbar_or} and~\eqref{eq: z_or} in their compact forms:
	\begin{subequations} 
		\label{eq: alg_conv}
		\begin{align}
			\tilde{\mathbf{w}}_{k+1} &= \mathbf{R}\tilde{\mathbf{w}}_{k} + \beta_{k} \mathbf{z}_{k}, \\
			\mathbf{z}_{k+1} &= \mathbf{C}\mathbf{z}_{k} - \iota (\mathbf{w}_{k+1} - \mathbf{w}_{k}).
		\end{align}
	\end{subequations}
	By setting $\mathbf{1}^{T}\mathbf{z}_{0} = - \iota \left(\sum_{i=1}^{N}w_{i,0}-\sum_{i=1}^{N}d_{i} \right)$, we can deduce using induction that
	\begin{equation*}
		\mathbf{1}^{T}\mathbf{z}_{k} = - \iota
		\left(\sum_{i=1}^{N} w_{i,k} - \sum_{i=1}^{N}d_{i} \right),
	\end{equation*}
	which implies that the agents can track the global mismatch between resource supply and demand.
	
	However, when exchanged messages are subject to noise, i.e., the received values are $\tilde{w}_{i,k} + \zeta_{i,k}$ and $z_{i,k} + \xi_{i,k}$ instead of $\tilde{w}_{i,k}$ and $z_{i,k}$, respectively, the update of the conventional algorithm~\eqref{eq: alg_conv} becomes 
	\begin{subequations}
		\begin{align}
			\tilde{\mathbf{w}}_{k+1} &= \mathbf{R}(\tilde{\mathbf{w}}_{k} + \boldsymbol{\zeta}_{k}) + \beta_{k} \mathbf{z}_{k}, \\
			\mathbf{z}_{k+1} &= \mathbf{C}(\mathbf{z}_{k} + \boldsymbol{\xi}_{k}) - \iota (\mathbf{w}_{k+1} - \mathbf{w}_{k}),
		\end{align}
	\end{subequations}
	where $\boldsymbol{\zeta}_{k} = \left[\zeta_{1,k}, \dots,\zeta_{N,k} \right]^{T} \in \mathbb{R}^{N \times m}$, $\boldsymbol{\xi}_{k} = \left[\xi_{k}, \dots, \xi_{N,k} \right]^{T} \in \mathbb{R}^{N \times m}$, and $\zeta_{i,k} \in \mathbb{R}^{m}$ and $\xi_{i,k} \in \mathbb{R}^{m}$ are injected noises.
	Using induction, we can deduce that even under $\mathbf{1}^{T}\mathbf{z}_{0} = -\iota \left(\sum_{i=1}^{N}w_{i,0}-\sum_{i=1}^{N}d_{i} \right)$:
	\begin{equation}
		\mathbf{1}^{T}\mathbf{z}_{k} = - \iota
		\left(\sum_{i=1}^{N} w_{i,k} - \sum_{i=1}^{N}d_{i} \right) + \sum_{l=0}^{k-1}\mathbf{1}^{T}\boldsymbol{\xi}_{l}.
	\end{equation}
    In the conventional algorithm, information-sharing noise accumulates over iterations, increasing total variance and significantly affecting optimization accuracy. To overcome the limitations of existing dual gradient-tracking algorithms~(\cite{zhang2020distributed, ding2021differentiallyra}) and reduce the impact of this noise, we draw inspiration from the robust push-pull method~(\cite{pu2020robust}). Instead of sharing the direct per-step global deviation estimate $z_{i,k}$, each agent shares the cumulative deviation estimate $s_{i,k}$. Our privacy-preserving method is outlined in Algorithm~\ref{algo: one}.
	\begin{algorithm}[t]
		\begin{algorithmic}[1] 
			\State Input: Step size sequence $\{\alpha_{k}\}$, noise sequences $\{ \xi_{i,k}\}$ and $\{\zeta_{i,k}\}$ for any $i \in \mathcal{N}$ and $k \geq 0$, and the parameters $\gamma$ and $\phi$.
			\State Initialization: $w_{i,0}, s_{i,0}, \tilde{w}_{i,0} \in \mathbb{R}^{m}$.
			\For {$k = 1, 2, \dots, $}
			\State for each $i \in \mathcal{N}$,\\
			\quad Agent $i$ pushes $C_{li}(s_{i,k} + \xi_{i,k})$ to each agent $l \in \mathcal{N}_{\mathbf{C}, i}^{\text{out}}$.\\
			\quad Agent $i$ pulls $\tilde{w}_{j,k} + \zeta_{j,k}$ from each agent $j \in \mathcal{N}_{\mathbf{R}, i}^{\text{in}}$.
			\vspace{1mm}
			\State for each $i \in \mathcal{N}$,
			\vspace*{-3mm}
			\begin{subequations} \label{eq: algo}
				\begin{align}
					\label{eq: tracking} s_{i,k+1} =& (1-\gamma)s_{i,k} + \gamma \sum_{j=1}^{N}C_{ij}\left( s_{j,k} + \xi_{j,k} \right) \nonumber \\ 
					&- \alpha_{k} (w_{i,k} - d_{i}), \\
					\label{eq: primal_update} w_{i,k+1} =& \mathop{\arg\min}\limits_{w \in \mathcal{W}_{i}} \left\{ F_{i}(w) - \tilde{w}_{i,k+1}^{T}w \right\}, \\
					\label{eq: dual_update} \tilde{w}_{i,k+1} =& (1-\phi)\tilde{w}_{i,k} + \phi \sum_{j=1}^{N}R_{ij}\left(\tilde{w}_{j,k} + \zeta_{j,k} \right) \nonumber  \\
					&+ (s_{i,k+1} - s_{i,k}).
				\end{align}
			\end{subequations}
			\EndFor
			\caption{Differentially Private Dual Gradient Tracking (DP-DGT)} \label{algo: one}
		\end{algorithmic}
	\end{algorithm}
	The Laplace mechanism is a fundamental technique for achieving DP and we thus assume that the noise satisfies Assumption~\ref{assum: noise}. Although Gaussian noise can also be employed, it may require a slight relaxation of the definition of DP~(\cite{dwork2006differential}).
	\begin{assum} \label{assum: noise}
		The noise $\xi_{i,k}$ and $\zeta_{i,k}$ are independently drawn by agent $i$ from the following zero-mean Laplace distribution,
		\begin{equation*}
			\xi_{i,k} \sim \emph{Lap}(\theta_{\xi, k}), \ \zeta_{i,k}  \sim \emph{Lap}(\theta_{\zeta, k}),
		\end{equation*}
		where $\{\theta_{\xi, k} \}$ and $\{\theta_{\zeta, k}\}$ are sequences to be designed.
	\end{assum}
	
	Defining $\mathbf{s}_{k} = [s_{1,k}, \dots, s_{N,k}]^{T} \in \mathbb{R}^{N \times m}$, we can rewrite~\eqref{eq: tracking} and~\eqref{eq: dual_update} from Algorithm~\ref{algo: one} as follows:
	\begin{subequations} \label{eq: alg_compact}
		\begin{align}
			\label{eq: s_compact} \mathbf{s}_{k+1} =& (1-\gamma) \mathbf{s}_{k} + \gamma \mathbf{C}(\mathbf{s}_{k} + \boldsymbol{\xi}_{k}) - \alpha_{k} (\mathbf{w}_{k}-\mathbf{d}), \\
			\tilde{\mathbf{w}}_{k+1} =& (1-\phi) \tilde{\mathbf{w}}_{k} + \phi \mathbf{R}(\tilde{\mathbf{w}}_{k} + \boldsymbol{\zeta}_{k}) +  (\mathbf{s}_{k+1} - \mathbf{s}_{k}).
		\end{align}
	\end{subequations} 
	In this compact form, we observe that $\mathbf{s}_{k+1} - \mathbf{s}_{k}$ is fed into the dual variable update and serves as the deviation estimate. 
	This approach prevents the accumulation of information noise on the global mismatch estimate. In fact, using the update rule of $\mathbf{s}_{k}$ in~\eqref{eq: s_compact}, and by letting $\mathbf{z}_{k} = \mathbf{s}_{k+1} - \mathbf{s}_{k} $, we obtain:
	\begin{equation}
		\begin{aligned}
			& \mathbf{1}^{T}\mathbf{z}_{k} =  \mathbf{1}^{T}(\mathbf{s}_{k+1} - \mathbf{s}_{k})\\
			=& -\gamma \mathbf{1}^{T}\mathbf{s}_{k} + \gamma \mathbf{1}^{T} \mathbf{C} (\mathbf{s}_{k} + \boldsymbol{\xi}_{k}) - \alpha_{k} \left( \sum_{i=1}^{N} w_{i,k} - \sum_{i=1}^{N} d_{i}  \right) \\
			=& - \alpha_{k} \left( \sum_{i=1}^{N} w_{i,k} - \sum_{i=1}^{N} d_{i}\right ) + \gamma \mathbf{1}^{T} \boldsymbol{\xi}_{k},
		\end{aligned}
	\end{equation}
	regardless of the initial selection of $\mathbf{s}_{0}$, where we used the property $\mathbf{1}^{T}\mathbf{C} = \mathbf{1}^{T}$ from Assumption~\ref{assum: rc_stochastic}. 
	Thus, the proposed algorithm utilizes $\mathbf{s}_{k+1} - \mathbf{s}_{k}$ to track the global deviation and prevent noise accumulation in the deviation tracking.
	Furthermore, in contrast to~\cite{zhang2020distributed, ding2021differentiallyra}, our algorithm does not have any requirements on the initialization.
    
	\section{Convergence Analysis} \label{sec: convergence}
By leveraging strong duality, we establish the convergence of Algorithm~\ref{algo: one} by proving the convergence of the dual problem~\eqref{eq: dual_form} under robust push-pull with information-sharing noise. However, existing convergence results for dual optimization (DO) with robust push-pull require the objective function $f_{i}$ to be strongly convex and Lipschitz smooth~(\cite{pu2020robust, chen2023differentially}). It is important to note that the dual objective $f_{i}$ in~\eqref{eq: dual_form} often loses strong convexity due to the convex conjugate function $F_{i}^{*}$, even when $F_i$ in~\eqref{eq: problem} is strongly convex. Therefore, we extend the convergence of dual variables under robust push-pull to non-convex objectives.
Additionally, \cite{huang2024differential} show that $\epsilon$-DP cannot be achieved with Laplace noise if step sizes are not summable (i.e., $\sum_{k=0}^{\infty} \alpha_{k} = \infty$ and $\sup_{k \geq 0} \alpha_{k} < \infty$). Thus, only summable step sizes ensure meaningful convergence and privacy performance under Laplacian noise.

We first demonstrate that, with the proposed algorithm and summable step sizes, dual variables for non-convex $f_{i}$ can converge to a neighborhood of a stationary point. Then, we show the convergence of primal variables in problem~\eqref{eq: problem} under DP-DGT.	
	
	\subsection{Convergence of Dual Variables}
	To better illustrate the difference of our convergence analysis with existing works using push-pull-based gradient-tracking methods, we rewrite the proposed algorithm~\eqref{eq: algo} in a typical push-pull form by letting $y_{i,k} = s_{i,k}$. Since $x_{i} = -\tilde{w}_{i}$, we have
	\begin{subequations} \label{eq: dual_algo}
		\begin{align}
			y_{i,k+1} =& (1-\gamma)y_{i,k} + \gamma \sum_{j=1}^{N}C_{ij} (y_{j,k} + \xi_{j,k}) + \alpha_{k} \nabla f_{i}(x_{i,k}), \\
			x_{i,k+1} =& (1-\phi)x_{i,k} + \phi \sum_{j=1}^{N}R_{ij}( x_{j,k} - \zeta_{j,k}) \nonumber \\
			&  - (y_{i,k+1} - y_{i,k}).
		\end{align}
	\end{subequations}
	Different from~\cite{ding2021differentially}, \cite{chen2023differentially}, and \cite{pu2020robust}, the $f_{i}, \forall i \in \mathcal{N}$ here is not strongly convex. Let $\mathbf{x}_{k} = -\tilde{\mathbf{w}}_{k}$ and $y_{k} = -\mathbf{s}_{k}$ and denote $G(\mathbf{x}) = \left[ \nabla f_{1}(x_{1,k}), \dots, \nabla f_{N}(x_{N,k}) \right]^{T} \in \mathbb{R}^{N \times m}$.
	Defining
	\begin{equation*}
		\begin{aligned}
			\mathbf{C}_{\gamma} &\coloneqq (1-\gamma)\mathbf{I} + \gamma \mathbf{C}, \\
			\mathbf{R}_{\phi} &\coloneqq (1-\phi)\mathbf{I} + \phi \mathbf{R},
		\end{aligned}
	\end{equation*}
	we can write~\eqref{eq: dual_algo} in the following compact form: 
	\begin{align}
		\label{eq: y_compact} \mathbf{y}_{k+1} =& \mathbf{C}_{\gamma} \mathbf{y}_{k} + \alpha_{k} G(\mathbf{x}_{k}) + \gamma \mathbf{C} \boldsymbol{\xi}_{k}, \\
		\label{eq: x_compact} \mathbf{x}_{k+1} =& \mathbf{R}_{\phi} \mathbf{x}_{k} + \phi \mathbf{R} \boldsymbol{\zeta}_{k} - (\mathbf{y}_{k+1} - \mathbf{y}_{k}) \nonumber \\
		=& \mathbf{R}_{\phi} \mathbf{x}_{k} -  \mathbf{v}_{k} + \phi \mathbf{R} \boldsymbol{\zeta}_{k} - \gamma  \mathbf{C} \boldsymbol{\xi}_{k},
	\end{align}
	where 
	we denote $\mathbf{v}_{k} = (\mathbf{C}_{\gamma} - I) \mathbf{y}_{k} + \alpha_{k} G(\mathbf{x}_{k})$. It can be verified that $\mathbf{R}_{\phi}$ and $\mathbf{C}_{\gamma}$ are row-stochastic and column-stochastic, respectively.
	Under Assumption~\ref{assum: spanning_tree}, we have some preliminary lemmas regarding the communication graphs.
	\begin{lemma} \label{lem: eigen}
		(\cite{horn2012matrix}) Suppose Assumption~\ref{assum: spanning_tree} holds.
		The matrix $\mathbf{R}$ has a unique unit nonnegative left eigenvector $\pi_{\mathbf{R}}$ w.r.t. eigenvalue $1$, i.e., $\pi_{\mathbf{R}}^{T}\mathbf{R} = \pi_{\mathbf{R}}^{T}$ and $\pi_{\mathbf{R}}^{T}\mathbf{1} = 1$. The matrix $\mathbf{C}$ has a unique unit nonnegative right eigenvector $\pi_{\mathbf{C}}$ w.r.t. eigenvalue $1$, i.e., $\mathbf{C}\pi_{\mathbf{C}} = \pi_{\mathbf{C}}$ and $\pi_{\mathbf{C}}^{T}\mathbf{1} = 1$.
	\end{lemma}
	Based on Lemma~\ref{lem: eigen} and the definition of $\mathbf{R}_{\phi}$ and $\mathbf{C}_{\gamma}$, we can also deduce that $\pi_{\mathbf{R}}^{T}\mathbf{R}_{\phi} = \pi_{\mathbf{R}}^{T}$ and $\mathbf{C}_{\gamma}\pi_{\mathbf{C}} = \pi_{\mathbf{C}}$.
	
	\begin{lemma} \label{lem: sigma}
		(\cite{pu2020push})
		Suppose Assumption~\ref{assum: spanning_tree} holds.
		There exist matrix norms $\left\| \cdot \right\|_{R}$ and $\left\| \cdot \right\|_{C}$ such that $\sigma_{R} := \left\| \mathbf{R}_{\phi} - \mathbf{1}\pi_{\mathbf{R}}^{T} \right\|_{R} < 1$ and $\sigma_{C} := \left\| \mathbf{C}_{\gamma} - \pi_{\mathbf{C}} \mathbf{1}^{T} \right\|_{C} < 1$. Furthermore, $\sigma_{R}$ and $\sigma_{C}$ can be arbitrarily closed to the spectral radius of $\mathbf{R}_{\phi} - \mathbf{1}\pi_{\mathbf{R}}^{T}$ and $\mathbf{C}_{\gamma} - \pi_{\mathbf{C}} \mathbf{1}^{T}$.
	\end{lemma}
	Note that the norms $\left\| \cdot \right\|_{R}$ and $\left\| \cdot \right\|_{C}$ are only for matrices, which is defined as 
	\begin{equation*}
		\left\| X \right\|_{R} = \| \tilde{R}X \tilde{R}^{-1} \|_{2} \ \text{and} \ \left\| X \right\|_{C} = \| \tilde{C}^{-1}X \tilde{C} \|_{2} 
	\end{equation*}
	for any matrix $X \in \mathbb{R}^{N \times N}$, where $\tilde{R}$ and $\tilde{C}$ are some invertible matrices. To facilitate presentation, we slightly abuse the notations and define vectors norm $\|x \|_{R} = \| \tilde{R}x \|_{2} $ and $\|x \|_{C} = \| \tilde{C}^{-1}x \|_{2} $ for any $x \in \mathbb{R}^{N}$.
	
	\begin{lemma} (\cite{pu2020push})
		There exist constants $\delta_{R,C}, \delta_{C,R}$ such that for all $\mathbf{x}$, we have $\left\| \cdot \right\|_{R} \leq \delta_{R,C} \left\| \cdot \right\|_{C}$ and $\left\| \cdot \right\|_{C} \leq \delta_{C,R} \left\| \cdot \right\|_{R}$. Additionally, we can easily obtain $\left\| \cdot \right\|_{R} \leq  \left\| \cdot \right\|_{2}$ and $\left\| \cdot \right\|_{C} \leq  \left\| \cdot \right\|_{2}$ from the construction of the norm $\left\| \cdot \right\|_{R}$ and $\left\| \cdot \right\|_{C}$.
	\end{lemma}
	
	Denote $\bar{\mathbf{x}}_{k} = \mathbf{x}_{k}^{T}\pi_{\mathbf{R}}$, 
	$\bar{\mathbf{v}}_{k} = \mathbf{v}_{k}^{T}\pi_{\mathbf{R}}$, 
	$\hat{\mathbf{v}}_{k} = \mathbf{v}_{k}^{T}\mathbf{1} = \alpha_{k} G(\mathbf{x}_{k})^{T}\mathbf{1}$, and let $\mathcal{F}_{k}$ be the $\sigma$-algebra generated by $\{ \boldsymbol{\xi}_{l}, \boldsymbol{\zeta}_{l} \}_{l = 0, \dots, k-1}$. With the above norms, we first establish a system of linear inequalities w.r.t. the expectations of $\left\| \mathbf{x}_{k+1} - \mathbf{1}_{N} \bar{\mathbf{x}}_{k+1}^{T}  \right\|^{2}_{R}$ and $\left\| \mathbf{v}_{k+1} - \pi_{\mathbf{C}}\hat{\mathbf{v}}_{k+1}^{T} \right\|_{\mathbf{C}}^{2}$ for the dual algorithm~\eqref{eq: dual_algo}.
	\begin{lemma} \label{lem: LMI} 
		Under Assumptions~\ref{assum: spanning_tree}--\ref{assum: noise} and the $L$-Lipschitz smoothness of $f_{i}, \forall i \in \mathcal{N}$, we have the following linear system of inequalities
		\begin{equation} \label{eq: LMI}
			\begin{aligned}
				&  \begin{bmatrix}
					\mathbb{E}\left[\left\| \mathbf{x}_{k+1} - \mathbf{1}_{N} \bar{\mathbf{x}}_{k+1}^{T}  \right\|^{2}_{R} \big| \mathcal{F}_{k} \right] \\
					\mathbb{E}\left[\left\| \mathbf{v}_{k+1} - \pi_{\mathbf{C}}\hat{\mathbf{v}}_{k+1}^{T} \right\|_{C}^{2} \big| \mathcal{F}_{k}  \right ] 
				\end{bmatrix}
				\\
				\preceq & 
				\underbrace{\begin{bmatrix}
						P_{11,k} & P_{12} \\ P_{21,k} & P_{22,k}
				\end{bmatrix}}_{\mathbf{P}_{k}}
				\begin{bmatrix}
					\left\| \mathbf{x}_{k} - \mathbf{1}_{N} \bar{\mathbf{x}}_{k}^{T}  \right\|^{2}_{R} \\
					\left\| \mathbf{v}_{k} - \pi_{\mathbf{C}}\hat{\mathbf{v}}_{k}^{T} \right\|_{C}^{2}
				\end{bmatrix} 
				\\
				+ &
				\underbrace{\begin{bmatrix}
						B_{11} & B_{12} & B_{13} \\ B_{21} & B_{22} & B_{23,k}
				\end{bmatrix}}_{\mathbf{B}_{k}}
				\begin{bmatrix}
					\alpha_{k}^{2}\mathbb{E}\left[\left\| \nabla f(\bar{\mathbf{x}}_{k})\right\|_{2}^{2} \big| \mathcal{F}_{k} \right] \\
					\phi^{2} \theta_{\zeta,k}^{2} \\ \gamma^{2} \theta_{\xi,k}^{2}
				\end{bmatrix},
			\end{aligned}
		\end{equation}
		where the inequality is taken element-wise, and the elements of matrices $\mathbf{P}_{k}$ and $\mathbf{B}_{k}$ are given by
		\vspace{-0.8cm}
		{ \small
		\begin{align*} 
			& P_{11,k} = \frac{1+\sigma_{R}^{2}}{2} + \mathbf{a}_{1} \alpha_{k}^{2}, \quad P_{12} = \mathbf{a}_{2}, \\
			&P_{21,k} = (\mathbf{a}_{3} + \mathbf{a}_{4} \alpha_{k}^{2})  \max\{ \alpha_{k+1}^{2}, \alpha_{k}^{2} \} , \\
			&P_{22,k} = \frac{1+\sigma_{C}^{2}}{2} + \mathbf{a}_{5}  \max\{ \alpha_{k+1}^{2}, \alpha_{k}^{2}  \},
		\end{align*} }
		and 
		\vspace{-1cm}
		{\small
		\begin{equation*}
			\begin{aligned}
				B_{11} &= 4 \frac{1+\sigma_{R}^{2}}{1-\sigma_{R}^{2}} \left\| I_{N} - \mathbf{1}\pi_{\mathbf{R}}^{T} \right\|_{R}^{2}  
				\left\| \pi_{\mathbf{C}} \right\|_{R}^{2}, \\
				B_{12} &= Nm \left\| \mathbf{R} - \mathbf{1}\pi_{\mathbf{R}}^{T} \right\|_{R}^{2},  \ B_{13} = Nm \left\| (I - \mathbf{1}\pi_{\mathbf{R}}^{T}) \mathbf{C} \right\|_{R}^{2}, \\
				B_{21} &= 2 \left\| \pi_{\mathbf{C}} \right\|_{C}^{2}\mathbf{a}_{4}, \ B_{22} = \frac{1}{3} \left\| \mathbf{R} \right\|_{C}^{2}Nm\mathbf{a}_{5}, \\
				B_{23,k} &= Nm\left[ \left\| 2(\mathbf{C}_{\gamma} -I)\mathbf{C} \right\|_{C}^{2} + 4L \alpha_{k+1} \right. \\
				& \quad  + \left. L^{2}\left\| \mathbf{C} \right\|_{C}  \max\{ \alpha_{k+1}^{2}, \alpha_{k}^{2} \}  \right] \mathbf{a}_{6},
			\end{aligned}
		\end{equation*} }
		with the constants in the following:
		\vspace{-1cm}
		{\small
		\begin{align*}
			\mathbf{a}_{1} =&  4 \frac{1+\sigma_{R}^{2}}{1-\sigma_{R}^{2}} \left\| I_{N} - \mathbf{1}\pi_{\mathbf{R}}^{T} \right\|_{R}^{2} \left\| \pi_{\mathbf{C}} \right\|_{R}^{2} L^{2} N,  \\
			\mathbf{a}_{2} =& 2\frac{1+\sigma_{R}^{2}}{1-\sigma_{R}^{2}} \left\| I_{N} - \mathbf{1}\pi_{\mathbf{R}}^{T} \right\|_{R}^{2} \delta_{R,C}^{2}, \\
			\mathbf{a}_{3} =& 3 \frac{1+\sigma_{C}^{2}}{1-\sigma_{C}^{2}} \left\| I - \pi_{\mathbf{C}}\mathbf{1}^{T} \right\|_{C}^{2} \left\| \mathbf{R}_{\phi}-I \right\|_{C}^{2} L^{2} \delta_{C,R}^{2}, \\
			\mathbf{a}_{4} =& 6 \frac{1+\sigma_{C}^{2}}{1-\sigma_{C}^{2}}\left\| I - \pi_{\mathbf{C}}\mathbf{1}^{T} \right\|_{C}^{2}L^{2} \delta_{C,R}^{2},  \\
			\mathbf{a}_{5} =& 3 \frac{1+\sigma_{C}^{2}}{1-\sigma_{C}^{2}} \left\| I - \pi_{\mathbf{C}}\mathbf{1}^{T} \right\|_{C}^{2} L^{2},  \\
			\mathbf{a}_{6} =& \frac{1+\sigma_{C}^{2}}{1-\sigma_{C}^{2}} \left\| I - \pi_{\mathbf{C}}\mathbf{1}^{T} \right\|_{C}^{2}.
		\end{align*}}
	\end{lemma}
	
\begin{pf}
		The proof is provided in Appendix A.2~(\cite{huo2024differentially}).
\end{pf}

	Note that when the step size $\alpha_{k} \to 0$, the matrix $\mathbf{P}_{k}$ tends to become upper-triangular. Its eigenvalues approach $q_{R} = \frac{1+\sigma_{R}^{2}}{2}$ and $q_{C} = \frac{1+\sigma_{C}^{2}}{2}$, where $\sigma_{R}<1$ and $\sigma_{C}<1$ are defined in Lemma~\ref{lem: sigma}. In the following, we establish conditions regarding step sizes and variances of Laplacian noise for the convergence of~\eqref{eq: dual_algo} with non-convex $f_{i}, \forall i \in \mathcal{N}$.
	
	\begin{theorem} \label{thm: convergence}
		Under Assumptions~\ref{assum: spanning_tree}--\ref{assum: noise} and the $L$-Lipschitz smoothness of $f_{i}, \forall i \in \mathcal{N}$, when $\sum_{k=0}^{\infty} \alpha_{k} < \infty$, $\sum_{k=0}^{\infty}\theta_{\xi,k}^{2} < \infty$, $\sum_{k=0}^{\infty}\theta_{\zeta,k}^{2} < \infty$, $\sum_{k=0}^{\infty} \frac{\theta_{\xi,k}^{2}}{\alpha_{k}}< \infty$, $\sum_{k=0}^{\infty} \frac{\theta_{\zeta,k}^{2}}{\alpha_{k}}< \infty$, and there exists $\lambda$ satisfying $q_{R} < \lambda < 1$ and $q_{C} < \lambda < 1$ and $k_{0}>0$ such that $\frac{\alpha_{k}}{\alpha_{k_0}} \geq \beta \lambda^{k-k_{0}}$, we have that 
		\begin{itemize}
			\item[\romannumeral1)] \label{stat: thm1_1} $ \lim\limits_{k \to \infty} \mathbb{E}[\|\mathbf{x}_{k} - \mathbf{1}_{N}\bar{\mathbf{x}}_{k}^{T} \|] = 0$.
			\item[\romannumeral2)] $\mathbb{E}[f(\bar{\mathbf{x}}_{k})] - f^{*}$ converges to a finite value almost surely.
		\end{itemize} 
	\end{theorem}
	\begin{pf}
			We first bound $\| \bar{\mathbf{v}}_{k} \|^{2}$. According to the definition of $\bar{\mathbf{v}}_{k}$, we have
		\begin{align*}
			\bar{\mathbf{v}}_{k} =&  ( \mathbf{v}_{k} -  \pi_{\mathbf{C}}\hat{\mathbf{v}}_{k}^{T} + \pi_{\mathbf{C}}\hat{\mathbf{v}}_{k}^{T})^{T}\pi_{\mathbf{R}} \\
			=& ( \mathbf{v}_{k} -  \pi_{\mathbf{C}}\hat{\mathbf{v}}_{k}^{T})^{T} \pi_{\mathbf{R}} 
			+ \alpha_{k} \left( G(\mathbf{x}_{k}) - G(\mathbf{1}\bar{\mathbf{x}}_{k}^{T}) \right)^{T} \mathbf{1} \pi_{\mathbf{C}}^{T}\pi_{\mathbf{R}} \\
			&+ \alpha_{k} \nabla f(\bar{\mathbf{x}}_{k})
			\pi_{\mathbf{C}}^{T}\pi_{\mathbf{R}},
		\end{align*}
		and hence, 
		\begin{align*}
			\left\| \bar{\mathbf{v}}_{k} \right\|^{2} \leq & 3 \left\| \pi_{\mathbf{R}} \right\|^{2} \delta_{2,C}^{2} \left\| \mathbf{v}_{k} -  \pi_{\mathbf{C}}\hat{\mathbf{v}}_{k}^{T} \right\|^{2}_{C} \\
			& + 3L^{2}N (\pi_{\mathbf{C}}^{T}\pi_{\mathbf{R}})^{2} \delta_{2,R}^{2} \alpha_{k}^{2} \left\| \mathbf{x}_{k} - \mathbf{1}\bar{\mathbf{x}}_{k}^{T} \right\|^{2}_{R} \\
			& + 3 (\pi_{\mathbf{C}}^{T}\pi_{\mathbf{R}})^{2} \alpha_{k}^{2} \left\| \nabla f(\bar{\mathbf{x}}_{k}) \right\|^{2}.
		\end{align*}
		For $- \nabla f(\bar{\mathbf{x}}_{k})^{T} \bar{\mathbf{v}}_{k}$, we have
		\begin{align*}
			& - \nabla f(\bar{\mathbf{x}}_{k})^{T} \bar{\mathbf{v}}_{k} \\
			\leq & - \pi_{\mathbf{C}}^{T}\pi_{\mathbf{R }} \alpha_{k} \left\| \nabla f(\bar{\mathbf{x}}_{k}) \right\|^{2} 
			+  \left\| \pi_{\mathbf{R}} \right\| \left\| \nabla f(\bar{\mathbf{x}}_{k}) \right\| \left\| \mathbf{v}_{k} -  \pi_{\mathbf{C}}\hat{\mathbf{v}}_{k}^{T} \right\| \\
			&+ L \sqrt{N} \pi_{\mathbf{C}}^{T}\pi_{\mathbf{R}} \alpha_{k}\left\| \nabla f(\bar{\mathbf{x}}_{k}) \right\| \left\| \mathbf{x}_{k} - \mathbf{1}\bar{\mathbf{x}}_{k}^{T} \right\|.
		\end{align*}	
		Since $f_{i}$ is $L$-Lipschitz smooth, we have $f(\bar{\mathbf{x}}_{k+1}) \leq f(\bar{\mathbf{x}}_{k}) + (\bar{\mathbf{x}}_{k+1} - \bar{\mathbf{x}}_{k})^{T}\nabla f(\bar{\mathbf{x}}_{k}) + \frac{L}{2} \left\| \bar{\mathbf{x}}_{k+1} - \bar{\mathbf{x}}_{k}\right\|^{2}$.
		According to $	\bar{\mathbf{x}}_{k+1} - \bar{\mathbf{x}}_{k} = (\mathbf{x}_{k+1} - \mathbf{x}_{k})^{T} \pi_{\mathbf{R}} 
		= - \mathbf{v}_{k}^{T}\pi_{\mathbf{R}} - \phi \boldsymbol{\zeta}_{k}^{T} \pi_{\mathbf{R}} - \gamma (\mathbf{C}\boldsymbol{\xi}_{k})^{T} \pi_{\mathbf{R}}$,
		we have 
		\begin{align} \label{eq: f_ave}
			&\mathbb{E}[f(\bar{\mathbf{x}}_{k+1})] \nonumber \\
			\leq & \mathbb{E}[f(\bar{\mathbf{x}}_{k})] - \mathbb{E}[\nabla f(\bar{\mathbf{x}}_{k})^{T}( \bar{\mathbf{v}}_{k} + \phi \boldsymbol{\zeta}_{k}^{T}\pi_{\mathbf{R}} + \gamma (\mathbf{C} \boldsymbol{\xi}_{k})^{T} \pi_{\mathbf{R}} )] \nonumber \\
			\leq & \mathbb{E}[f(\bar{\mathbf{x}}_{k})] + \frac{L}{2} \left\| \pi_{\mathbf{R}} \right\|^{2} \phi^{2} Nm \theta_{\zeta,k}^{2} \nonumber \\
			& + \frac{L}{2} \left\| \mathbf{C}^{T} \pi_{\mathbf{C}} \right\|^{2} \gamma^{2} Nm \theta_{\xi,k}^{2} \nonumber \\
			&+ \frac{\delta_{2,R}^{2}}{2}(\pi_{\mathbf{C}}^{T}\pi_{\mathbf{R}})^{2}(3L^{3} N  \alpha_{k}^{2}  + 1) \left\| \mathbf{x}_{k} - \mathbf{1}\bar{\mathbf{x}}_{k}^{T} \right\|^{2}_{R} \nonumber \\
			&+ \frac{\delta_{2,C}^{2}}{2 \alpha_{k} \pi_{\mathbf{C}}^{T}\pi_{\mathbf{R}}}\left\| \pi_{\mathbf{R}} \right\|^{2}(3L + 1) \left\| \mathbf{v}_{k} -  \pi_{\mathbf{C}}\hat{\mathbf{v}}_{k}^{T} \right\|^{2}_{C} \nonumber \\
			&- \bigg(  \frac{1}{2}\pi_{\mathbf{C}}^{T}\pi_{\mathbf{R}} \alpha_{k} \nonumber \\
			& \quad \quad - \frac{3L(\pi_{\mathbf{C}}^{T}\pi_{\mathbf{R}})^{2}  + L^{2}N}{2} \alpha_{k}^{2} \bigg) \mathbb{E}[\left\| \nabla f(\bar{\mathbf{x}}_{k}) \right\|^{2}].
		\end{align}	
		Since $\sum_{k=0}^{ \infty } \alpha_{k} < \infty$, there exists $K$ such that $\forall k \geq K$, one has $\alpha_{k} < \frac{\pi_{\mathbf{C}}^{T}\pi_{\mathbf{R}}}{3L(\pi_{\mathbf{C}}^{T}\pi_{\mathbf{R}})^{2} +1+L^{2}N }$.
		For $k \leq K$, there always exists a bound for $\mathbb{E}[\left\| \nabla f(\bar{\mathbf{x}}_{k}) \right\|^{2}]$, $\mathbb{E}\left[ \| \mathbf{x}_{k} - \mathbf{1}\bar{\mathbf{x}}_{k}^{T} \|^{2}_{R} \right]$, and $\mathbb{E}\left[ \| \mathbf{v}_{k} -  \pi_{\mathbf{C}}\hat{\mathbf{v}}_{k}^{T} \|^{2}_{C}  \right]$. Then we only need to prove the boundedness of these three values for $k > K$. 
		Specifically, denoting $g_{k} = \mathbb{E}[\left\| \nabla f(\bar{\mathbf{x}}_{k}) \right\|^{2}]$, $X_{k}= \mathbb{E}\left[ \| \mathbf{x}_{k} - \mathbf{1}\bar{\mathbf{x}}_{k}^{T} \|^{2}_{R}   \right]$, and $V_{k} = \mathbb{E}\left[ \| \mathbf{v}_{k} -  \pi_{\mathbf{C}}\hat{\mathbf{v}}_{k}^{T} \|^{2}_{C}  \right]$ when $k \geq K$. We will prove the following:
		\begin{equation} \label{eq: boundedness}
			g_{k} \leq D_{g}, \ X_{k} \leq D_{X}, \ V_{k} \leq D_{V}, \ \forall k > K,
		\end{equation}
		where $D_{g}, D_{X}, D_{V} >0$ are some constants.
		We prove~\eqref{eq: boundedness} by induction. Assume that~\eqref{eq: boundedness} holds form certain $k>0$, then we need to prove that
		\begin{subequations}
			\begin{align}
				\label{eq: Xk} X_{k+1} \leq & q_{R}X_{k} + \mathbf{a}_{1} \alpha_{k}^{2} D_{X} + \mathbf{a}_{2} V_{k} + B_{11} \alpha_{k}^{2} D_{g} \nonumber \\
				& + B_{12} \phi^{2} \theta_{\zeta,k}^{2} + B_{13} \gamma^{2} \theta_{\xi,k}^{2}, \\
				\label{eq: Vk} V_{k+1} \leq & q_{C}V_{k} + r_{1,k},
			\end{align}
		\end{subequations}
		with $r_{1,k} = \mathbf{a}_{5} \max\{ \alpha_{k+1}^{2}, \alpha_{k}^{2} \}D_{V}+ (\mathbf{a}_{3} \\ + \mathbf{a}_{4} \alpha_{k}^{2}) \max\{ \alpha_{k+1}^{2}, \alpha_{k}^{2}\} D_{x} + B_{21} \alpha_{k}^{2}D_{g} + B_{22}\phi^{2} \theta_{\zeta,k}^{2} + B_{23} \gamma^{2} \theta_{\xi,k}^{2}$, and then~\eqref{eq: Vk} suffices to show $V_{k+1} \leq q_{C}^{k+1} V_{0} + \sum_{l=0}^{k} q_{C}^{k-l}r_{1,l} \leq D_{V}$. 
		Since $\sum_{l=0}^{k} r_{1,l} \leq \sum_{l=0}^{\infty} r_{1,l}$ and
		\begin{align*}
			\sum_{l=0}^{\infty} r_{1,l} \leq &  (\mathbf{a}_{5} D_{V} + B_{21} D_{g} + \mathbf{a}_{3}D_{X}^{2}) \sum_{l=0}^{\infty} \alpha_{l}^{2} \\
            &+ B_{22} \phi^{2} \sum_{l=0}^{\infty} \theta_{\zeta,k}^{2} + B_{23}\gamma^{2} \sum_{l=0}^{\infty} \theta_{\xi,k}^{2}  \\
			&+ o(\sum_{l=0}^{\infty} \alpha_{l}^{2} + \sum_{l=0}^{\infty} \theta_{\xi,k}^{2} + \sum_{l=0}^{\infty} \theta_{\zeta,k}^{2}) \\
			=& D_{V}^{\prime} < \infty,
		\end{align*}
		by defining $D_{V} = V_{0} + D_{V}^{\prime}$, $V_{0}$ always satisfies $V_{0} \leq D_{V}$. By induction and $q_{C}<1$, we have $V_{k+1} < V_{0} + \sum_{l=0}^{k}r_{1,l} \leq D_{V}$.
		
Since $\frac{\alpha_{k}}{\alpha_{k_{0}}} \leq \beta \lambda^{k-k_{0}}$, one further has $V_{k+1} \leq q_{C}^{k+1} V_{0} + \sum_{l=0}^{k}q_{C}^{k-l}r_{1,l} \leq \left( \frac{q_{C}}{\lambda} \right)^{k+1} \alpha_{k+1} \frac{V_{0}}{\beta \alpha_{0}} + \frac{\alpha_{k+1}}{\beta \lambda} \sum_{l=0}^{k} \left( \frac{q_{C}}{\lambda} \right)^{k-l} \frac{r_{1,l}}{\alpha_{l}}$, which further yields that
\begin{align} \label{eq: Vk_alpha}
	\sum_{k=0}^{\infty} \frac{V_{k}}{\alpha_{k}} \leq & \frac{\lambda}{\lambda - q_{C}} \frac{V_{0}}{\beta \alpha_{0}} + \frac{1}{\beta(\lambda - q_{C})} \sum_{k=0}^{\infty}\frac{r_{1,k}}{\alpha_{k}} \nonumber \\
	\leq & \frac{\lambda}{\lambda - q_{C}} \frac{V_{0}}{\beta \alpha_{0}} + \frac{2 \mathbf{a}_{5} D_{V}}{\beta(\lambda - q_{C})} \sum_{l=0}^{\infty} \alpha_{l} \nonumber\\&+ \frac{B_{22} \phi^{2}}{\beta(\lambda - q_{C})} \sum_{l=0}^{\infty} \frac{\theta_{\zeta,k}^{2}}{\alpha_{k}^{2}} \nonumber \\
	&+ \frac{Nm \| 2(\mathbf{C}_{\gamma} - I) \mathbf{C} \|_{C}^{2} \gamma^{2}}{\beta(\lambda - q_{C})} \sum_{l=0}^{\infty} \frac{ \theta_{\xi,k}^{2}}{\alpha_{k}^{2}} \nonumber \\
	& + o\left(\sum_{l=0}^{\infty} \alpha_{l} + \sum_{l=0}^{\infty} \frac{\theta_{\xi,k}^{2}}{\alpha_{k}} \right) < \infty.
\end{align}
Therefore, we can infer $\sum_{k=0}^{\infty} V_{k} < \infty$. 
		
		Define $r_{2,k} = \mathbf{a}_{1} \alpha_{k}^{2} D_{X} + \mathbf{a}_{2} V_{k} + B_{11} \alpha_{k}^{2} D_{g} + B_{12} \phi^{2} \theta_{\zeta,k}^{2} + B_{13} \gamma^{2} \theta_{\xi,k}^{2}$, and then~\eqref{eq: Xk} suffices to show $X_{k+1} \leq q_{R}^{k+1} X_{0} + \sum_{l=0}^{k}q_{R}^{k-l}r_{2,l} \leq D_{X}$. We have
		\begin{align*}
			\sum_{k=0}^{\infty} r_{2,k} \leq & \sum_{k=0}^{\infty} (\mathbf{a}_{1}D_{X} + B_{11}D_{g}) \alpha_{k}^{2} + \mathbf{a}_{2} \sum_{k=0}^{\infty} V_{k} \\ 
			&+ B_{12} \phi^{2} \sum_{k=0}^{\infty} \theta_{\zeta,k}^{2} + B_{13} \gamma^{2} \sum_{k=0}^{\infty} \theta_{\xi,k}^{2} \\
			=& D_{X}^{\prime} < \infty.
		\end{align*} 
		By letting $D_{X} = X_{0} + D_{X}^{\prime}$, we have $X_{k+1} < X_{0} + \sum_{l=0}^{k}r_{2,l} \leq D_{X}$. 
		
		Similar to~\eqref{eq: Vk_alpha}, one has 
		\begin{align}\label{eq: Xk_alpha}
			& \sum_{k=0}^{\infty} \frac{X_{k}}{\alpha_{k}} \leq  \frac{\lambda}{\lambda - q_{R}} \frac{X_{0}}{\beta \alpha_{0}} + \frac{1}{\beta(\lambda - q_{R})} \sum_{k=0}^{\infty}\frac{r_{2,k}}{\alpha_{k}} \nonumber \\
			\leq & \frac{\lambda}{\lambda - q_{R}} \frac{X_{0}}{\beta \alpha_{0}} + \frac{\mathbf{a}_{1} D_{X} + B_{11} D_{g}}{\lambda - q_{R}} \sum_{k=0}^{\infty} \alpha_{k} + \frac{\mathbf{a}_{2}}{\lambda - q_{R}} \sum_{k=0}^{\infty}\frac{V_{k}}{\alpha_{k}} \nonumber \\
			& + \frac{B_{12} \phi^{2}}{\lambda - q_{R}} \sum_{k=0}^{\infty} \frac{\theta_{\zeta,k}^{2}}{\alpha_{k}} + \frac{B_{13} \gamma^{2}}{\lambda - q_{R}} \sum_{k=0}^{\infty} \frac{\theta_{\xi,k}^{2}}{\alpha_{k}} < \infty.
		\end{align}
		Hence, we have $\sum_{k=0}^{\infty} X_{k} < \infty$, and $\lim_{k \to \infty} X_{k} = 0$ since $X_{k} \geq 0$. 
		
		According to~\eqref{eq: f_ave}, we have 
		\begin{align} \label{eq: final_con}
			&\mathbb{E}[f(\bar{\mathbf{x}}_{k+1})] -f^{*} \leq  \mathbb{E}[f(\bar{\mathbf{x}}_{k})] - f^{*} \nonumber \\
			&- \left(\alpha_{k} \pi_{\mathbf{C}}^{T}\pi_{\mathbf{R}} - \frac{3L(\pi_{\mathbf{C}}^{T}\pi_{\mathbf{R}})^{2} + 1 + L^{2}N}{2} \alpha_{k}^{2} \right) \mathbb{E}[\left\| \nabla f(\bar{\mathbf{x}}_{k}) \right\|^{2}] \nonumber \\
			& + r_{3,k},
		\end{align}
		where $r_{3,k} = \frac{L}{2} \| \pi_{\mathbf{R}} \|^{2} \phi^{2} Nm \theta_{\zeta,k}^{2} + \frac{L}{2} \left\| \mathbf{C}^{T} \pi_{\mathbf{C}} \right\|^{2} \gamma^{2} Nm \theta_{\xi,k}^{2} \\ + \frac{\delta_{2,R}^{2}}{2}(\pi_{\mathbf{C}}^{T}\pi_{\mathbf{R}})^{2}(3L^{3} N  \alpha_{k}^{2}  + 1) X_{k} + \frac{\delta_{2,C}^{2}}{2}\left\| \pi_{\mathbf{R}} \right\|^{2}(3L + 1) \frac{V_{k}}{\alpha_{k}}$. Since $\sum_{k=0}^{\infty} \theta_{\zeta,k}^{2} < \infty$, $\sum_{k=0}^{\infty} \theta_{\xi,k}^{2} < \infty$, $\sum_{k=0}^{\infty} \frac{V_{k}}{\alpha_{k}} < \infty$, and $\sum_{k=0}^{\infty} X_{k} < \infty$,
		we have $\sum_{k=0}^{\infty} r_{3,k} < \infty$. 
		Furthermore, $\alpha_{k} < \frac{2\pi_{\mathbf{C}}^{T}\pi_{\mathbf{R}}}{3L(\pi_{\mathbf{C}}^{T}\pi_{\mathbf{R}})^{2} + 1 + L^{2}N}$ yields the non-negativeness of the third term in~\eqref{eq: final_con}.
		
		Based on Lemma 5 in Appendix A.1~(\cite{huo2024differentially}), we conclude that $\mathbb{E}[f(\bar{\mathbf{x}}_{k})] - f^{*}$ converges a.s. to a finite value and 
        \vspace{-8mm}
        {\small
        \begin{equation*}
        \begin{aligned}
           & \sum_{k = k^{\prime}}^{\infty} \big(\alpha_{k} \pi_{\mathbf{C}}^{T}\pi_{\mathbf{R}} - \frac{3L(\pi_{\mathbf{C}}^{T}\pi_{\mathbf{R}})^{2} + 1 + L^{2}N}{2} \alpha_{k}^{2} \big) \mathbb{E}[\left\| \nabla f(\bar{\mathbf{x}}_{k}) \right\|^{2}] \\
            &<  \infty.
        \end{aligned}
        \end{equation*}}
  
        Therefore, we conclude that $\sup_{k} \mathbb{E}[\left\| \nabla f(\bar{\mathbf{x}}_{k}) \right\|^{2}]$ exists, and thus we can define $D_{g} = \sup_{k} \mathbb{E}[\left\| \nabla f(\bar{\mathbf{x}}_{k}) \right\|^{2}]$.
	Here, we complete the proof of~\eqref{eq: boundedness} and additionally prove that $\lim_{k \to \infty} \mathbb{E}\left[ \| \mathbf{x}_{k} - \mathbf{1}\bar{\mathbf{x}}_{k}^{T} \|^{2}_{R}   \right] = 0$, and $\mathbb{E}[f(\bar{\mathbf{x}}_{k})] - f^{*}$ converges a.s. to a finite value.
	\end{pf}
    \begin{remark}      
   As shown in \eqref{eq: Vk_alpha}, \eqref{eq: Xk_alpha}, and \eqref{eq: final_con}, the convergence rate of $\mathbb{E}[f(\bar{\mathbf{x}}_k)] - f^*$ depends on the decay rates of $\sum_{k=0}^\infty \alpha_k$, $\sum_{k=0}^\infty \theta_{\xi,k}^2$, $\sum_{k=0}^\infty \theta_{\zeta,k}^2$, $\sum_{k=0}^\infty \frac{\theta_{\xi,k}^2}{\alpha_k}$, and $\sum_{k=0}^\infty \frac{\theta_{\zeta,k}^2}{\alpha_k}$ under $\sup_k \alpha_k < 1$. For example, choosing  parameters ($\alpha_k,\theta_{\xi,k},\theta_{\zeta,k}$) as $O(1/k^{1+p})$ for any $p>0$ yields an $O(1/k^p)$ convergence rate for DP-DGT.
    \end{remark}

	\begin{remark}
Theorem~\ref{thm: convergence} depends only on the smoothness of $f_i$, not its convexity, extending the analysis in \cite{pu2020robust, ding2021differentially, chen2023differentially, wang2023tailoring, huang2024differential}. This allows the method to apply to distributed non-convex optimization with information-sharing noise. Additionally, \cite[Theorem 2]{huang2024differential} shows that the gradient tracking algorithm with robust push-pull in~\eqref{eq: dual_algo} cannot achieve $\epsilon$-differential privacy if step sizes are not summable under Laplacian noise. Therefore, the convergence analysis with constant step sizes~(\cite{zhang2020distributed}) is inapplicable. We provide a rigorous convergence analysis for summable step size sequences under Laplace noise, subject to specific conditions.
	\end{remark}
	
	\subsection{Convergence of Primal Variables}
Since $f_i$ in~\eqref{eq: dual_form} is convex, Theorem~\ref{thm: convergence} ensures that under DP-DGT, the dual variables converge to a neighborhood of the optimal solution of~\eqref{eq: dual_form}. Thus, we can now establish the convergence of DP-DGT for the distributed resource allocation problem~\eqref{eq: problem}.
	\begin{theorem} \label{thm: primal_con}
		Under Assumptions~\ref{assum: convexity_slater}--\ref{assum: noise}, if $\sum_{k=0}^{\infty}\theta_{\xi,k}^{2} < \infty$, $\sum_{k=0}^{\infty}\theta_{\zeta,k}^{2} < \infty$, $\sum_{k=0}^{\infty} \alpha_{k} < \infty$, $\sum_{k=0}^{\infty} \frac{\theta_{\xi,k}^{2}}{\alpha_{k}}< \infty$, $\sum_{k=0}^{\infty} \frac{\theta_{\zeta,k}^{2}}{\alpha_{k}}< \infty$, and there exists $\lambda$ satisfying $q_{R} < \lambda < 1$ and $q_{C} < \lambda < 1$ and $k_{0}>0$ such that $\frac{\alpha_{k}}{\alpha_{k_0}} \geq \beta \lambda^{k-k_{0}}$,
		then the sequence $\{ \mathbf{w}_{k} \}$ in Algorithm~\ref{algo: one} converges to a neighborhood of $\mathbf{w}^{*}$, where $\mathbf{w}_{k} = \left[ w_{1,k}, \dots, w_{N,k} \right]^{T} \in \mathbb{R}^{N \times m}$ and $\mathbf{w}^{*} = \left[ w_{1}^{*}, \dots, w_{N}^{*} \right]^{T} \in \mathbb{R}^{N \times m}$.
	\end{theorem}
	\begin{pf}
		The proof is provided in Appendix A.3~(\cite{huo2024differentially}).
	\end{pf}

Our proposed algorithm exhibits constraint violations due to inherent errors in the dual variable solution, as shown in Theorem~\ref{thm: convergence}. This issue is common in works using Laplacian noise, as DP noise can affect convergence accuracy and constraint satisfaction. While some methods achieve accurate convergence~(\cite{munoz2021private, wang2024robust}), they are limited to specific problem types or privacy aspects. In conclusion, constrained distributed optimization faces challenges in balancing DP and constraint satisfaction, and we aim to develop advanced algorithms to address these violations in the future.

	\section{Differential Privacy Analysis} \label{sec: DP}
	In our analysis, we consider the worst-case scenario where the adversary can observe all communication in the network and has access to the initial value of the algorithm. 
	Thus, we denote the attacker's observation sequence as $\{\mathcal{O}_{k}\}_{k \geq 0}$, and the observation at time $k$ is $\mathcal{O}_{k} = \{ s_{i,k} + \xi_{i,k}, \tilde{w}_{i,k} + \zeta_{i,k} | \ \forall i \in \mathcal{N}  \}$.
	
	\begin{theorem} \label{thm: DP}
		Under Assumptions~\ref{assum: convexity_slater}--\ref{assum: noise}, if $\sum_{k=0}^{\infty} \alpha_{k} < \infty$, $D_{\alpha, \xi} := \sum_{k=0}^{\infty} \frac{\alpha_{k}}{\theta_{\xi,k}} < \infty$, $D_{\alpha, \zeta} := \sum_{k=0}^{\infty} \frac{\alpha_{k}}{\theta_{\zeta,k}} < \infty$, and $\pi_{\textbf{C}}^{T}\pi_{\textbf{R}} < \frac{1}{2}$, then Algorithm~\ref{algo: one} achieves $\epsilon$-differential privacy for any two $\delta$-adjacent distributed resource allocation problems, with the cumulative privacy budget
		\begin{equation}
			\epsilon = \frac{\delta + D_{\eta}}{\mu \gamma \phi} (D_{\alpha, \xi} + \phi D_{\alpha, \zeta}),
		\end{equation}
		where
		\begin{equation}
			D_{\eta}  = \inf_{K \geq k_{0}} \max \left\{ \frac{\max \limits_{0\leq l < K}\{ \alpha_{l}\delta + \alpha_{l}\eta_{l} \} + \bar{\alpha}\delta}{\gamma \phi \mu - \bar{\alpha}}, \max_{0 \leq l < K} \eta_{l}  \right\},
		\end{equation}
		with 
		\begin{equation*}
			\bar{\alpha} = \sup_{k} \alpha_{k} \ \text{and} \ k_{0} = \min_{k}\{ \alpha_{t} < \mu\gamma\phi \}.
		\end{equation*}
	\end{theorem}
\vspace{-0.5cm}
	\begin{pf}
		We consider the implementation of the proposed algorithm for both resource allocation problem $\mathcal{P}$ and $\mathcal{P}^{\prime}$. Since it is assumed that the attacker knows all auxiliary information, including the initial states, local demands, and the network topology, we have $\mathbf{s}_{0} = \mathbf{s}_{0}^{\prime}$, $\tilde{\mathbf{w}}_{0} = \tilde{\mathbf{w}}_{0}^{\prime}$, and $\mathbf{w}_{0} = \mathbf{w}^{\prime}_{0}$. 
		From Algorithm~\ref{algo: one}, it can be seen that given initial state $\{\mathbf{s}_{i, 0}, \tilde{\mathbf{w}}_{0}, \mathbf{w}_{0} \}$, the communication graphs $\{\mathbf{R}, \mathbf{C}\}$ and the function set, the observation sequence $\mathcal{O} = \{ \mathcal{O}_{k}\}$ is uniquely determined by the noise sequences $\{ \boldsymbol{\xi}_{k} \}$ and $\{ \boldsymbol{\zeta}_{k} \}$. 
		Thus, it is equivalent to prove that 
		\begin{equation*}
			\mathbb{P}[ \mathcal{R}^{-1}(\mathcal{P}, \mathcal{O}, \mathbf{s}_{0}, \tilde{\mathbf{w}}_{0}, \mathbf{w}_{0}) ] \leq e^{\epsilon} \mathbb{P}[ \mathcal{R}^{-1}(\mathcal{P}^{\prime}, \mathcal{O}, \mathbf{s}_{0}, \tilde{\mathbf{w}}_{0}, \mathbf{w}_{0}) ].
		\end{equation*}
		The attacker can eavesdrop on the transmitted messages, and therefore, there is $\mathbf{s}_{k} + \boldsymbol{\xi}_{k} = \mathbf{s}^{\prime}_{k}+\boldsymbol{\xi}^{\prime}_{k}$, and $\tilde{\mathbf{w}}_{k} + \boldsymbol{\zeta}_{k} = \tilde{\mathbf{w}}^{\prime}_{k} + \boldsymbol{\zeta}^{\prime}_{k}$.

		For any $i \neq i_{0}$, since $s_{i, 0} = s_{i,0}^{\prime}$, $w_{i,0} = w_{i,0}^{\prime}$, and $s_{j,0} + \xi_{j,0} = s_{j,0}^{\prime} + \xi_{j,0}^{\prime}$ for any $j \in \mathcal{N}$, we obtain $s_{i,1} = s_{i,1}^{\prime}$ based on~\eqref{eq: tracking}. Due to $\tilde{w}_{i,0} = \tilde{w}_{i,0}^{\prime}$ and $\tilde{w}_{j,0} + \zeta_{j,0} = \tilde{w}_{i,0}^{\prime} + \zeta_{j,0}^{\prime}$ for any $j \in \mathcal{N}$, it then follows that $\tilde{w}_{i,1} = \tilde{w}_{i, 1}^{\prime}$ according to~\eqref{eq: dual_update}. Also, $w_{i,1} = w_{i,1}^{\prime}$ can be inferred from~\eqref{eq: primal_update} due to $F_{i} = F_{i}^{\prime}$. Based on the above analysis, we can finally obtain
		\begin{equation} \label{eq: equal_noise}
			\xi_{i,k} = \xi_{i,k}^{\prime}, \ \zeta_{i,k} = \zeta_{i,k}^{\prime}, \ \forall k \geq 0, \ \forall i \neq i_{0}.
		\end{equation}
		For agent $i_{0}$, the noise should satisfy
		\begin{equation} \label{eq: noise}
			\Delta \xi_{i_{0},k} = -\Delta s_{i_{0},k}, \ \Delta \zeta_{i_{0},k} = -\Delta \tilde{w}_{i_{0},k}, \ \forall k >0,
		\end{equation}	
		where $\Delta \xi_{i_{0},k} = \xi_{i_{0},k} - \xi_{i_{0},k}^{\prime}$, $\Delta s_{i_{0},k} = s_{i_{0},k} - s_{i_{0},k}^{\prime}$, $\Delta \zeta_{i_{0},k} = \zeta_{i_{0},k} - \zeta_{i_{0},k}^{\prime}$, and $\Delta \tilde{w}_{i_{0},k} = \tilde{w}_{i_{0},k} - \tilde{w}_{i_{0},k}^{\prime}$. We have
		\begin{equation} \label{eq: sensitivity}
			\begin{aligned} 
				\Delta s_{i_{0},k+1} =& (1-\gamma) \Delta s_{i_{0},k} - \alpha_{k} \Delta w_{i_{0},k}, \\
				\Delta \tilde{w}_{i_{0},k+1} =& (1-\phi) \Delta \tilde{w}_{i_{0},k} + (\Delta s_{i_{0},k+1} - 	\Delta s_{i_{0},k}),
			\end{aligned}
		\end{equation}
		where $\Delta w_{i_{0},k} = w_{i_{0},k} - w_{i_{0},k}^{\prime}$. Based on the primal variable update~\eqref{eq: primal_update}, the following relationship holds:
		\begin{align*}
			&\| \tilde{w}_{i_{0},k} - \tilde{w}_{i_{0},k}^{\prime} \|_{1} 
			= \| \nabla F_{i_{0}}(w_{i_{0},k}) - \nabla F_{i_{0}}^{\prime}(w_{i_{0},k}^{\prime}) \|_{1} \nonumber \\
			=& \| \nabla F_{i_{0}}(w_{i_{0},k}) - \nabla F_{i_{0}}(w_{i_{0},k}^{\prime}) \\
			& \quad + \nabla F_{i_{0}}(w_{i_{0},k}^{\prime}) - \nabla F_{i_{0}}^{\prime}(w_{i_{0},k}^{\prime})  \|_{1} \nonumber \\
			\geq & \| \nabla F_{i_{0}}(w_{i_{0},k}) - \nabla F_{i_{0}}(w_{i_{0},k}^{\prime})  \|_{1} - \delta \nonumber \\
			\geq & \mu \| \Delta w_{i_{0},k} \|_{1} - \delta,
		\end{align*}
		where the first inequality is from Definition~\ref{defn: adjacency}.
		Therefore, we have
		\begin{equation}
			\| \Delta w_{i_{0},k} \|_{1} \leq \frac{\| \Delta \tilde{w}_{i_{0},k} \|_{1}}{\mu} + \frac{\delta}{\mu}.
		\end{equation}
		Then, taking the $\ell_{1}$-norm of both side of~\eqref{eq: sensitivity} yields
		\begin{equation}
			\begin{aligned}
				\| \Delta s_{i_{0},k+1} \|_{1} \leq & (1-\gamma)\| \Delta s_{i_{0},k} \|_{1} + \frac{\alpha_{k}}{\mu} \| \Delta \tilde{w}_{i_{0},k} \|_{1} + \frac{\alpha_{k} \delta}{\mu}, \\
				\| \Delta \tilde{w}_{i_{0},k+1} \|_{1} \leq & \left(1-\phi + \frac{\alpha_{k}}{\mu } \right) \| \Delta \tilde{w}_{i_{0},k} \|_{1} \nonumber \\
				&+ (2- \gamma) \| \Delta s_{i_{0},k} \|_{1} + \frac{\alpha_{k} \delta}{\mu}.
			\end{aligned}
		\end{equation}
		Consider a discrete-time dynamical system,
		\begin{equation} \label{eq: DTD}
			\begin{bmatrix}
				\varphi_{k+1} \\ \eta_{k+1}
			\end{bmatrix}
			= 
			\begin{bmatrix}
				1-\gamma & \frac{\alpha_{k}}{\mu } \\
				2-\gamma & 1-\phi + \frac{\alpha_{k}}{\mu }
			\end{bmatrix}
			\begin{bmatrix}
				\varphi_{k} \\ \eta_{k}
			\end{bmatrix}
			+ \frac{\alpha_{k} \delta}{\mu}
			\begin{bmatrix}
				1 \\ 1
			\end{bmatrix},
		\end{equation}
		with $\varphi_{0} = 0$ and $\eta_{0} = 0$. 
		Since $\| \Delta s_{i_{0},0} \| \leq \varphi_{0}$ and $\| \Delta \tilde{w}_{i_{0},0} \|  \leq \eta_{0}$, we infer that $\| \Delta s_{i_{0},k} \|_{1}\leq \varphi_{k}$ and $\| \Delta \tilde{w}_{i_{0},k} \|_{1} \leq \eta_{k}$, $\forall k \geq 0$, by induction.
		Moreover, we have $\begin{bmatrix}
			\varphi_{k+1} \\ \eta_{k+1}
		\end{bmatrix} = \sum_{l=0}^{k} \left (\frac{\alpha_{l} \delta}{\mu} + \frac{\alpha_{l} \eta_{l} }{\mu } \right ) P^{k-l} \begin{bmatrix}
			1 \\ 1
		\end{bmatrix} $, with $P^{0} = I$ and $P^{k} = \begin{bmatrix}
			(1-\gamma)^{k} & 0 \\ \sum_{l=0}^{k-1}(2-\gamma)(1-\phi)^{k-1-l} (1-\gamma)^{l} & (1-\phi)^{k}
		\end{bmatrix}$ for $k > 0$. 
		Hence, 
		\begin{align} 
			\varphi_{k+1} =& \sum_{l=0}^{k} (1-\gamma)^{k-l}\left( \frac{\alpha_{l} \delta}{\mu} + \frac{\alpha_{l} \eta_{l} }{\mu } \right), \\
			\label{eq: eta} \eta_{k+1} 
			=& \sum_{l=0}^{k}\Bigg[ \sum_{j=0}^{k-l-1} (2-\gamma)(1-\phi)^{k-1} \left(\frac{1-\gamma}{1-\phi} \right)^{l} \nonumber \\
			& + (1-\phi)^{k-l} \Bigg] \left( \frac{\alpha_{l} \delta}{\mu} + \frac{\alpha_{l} \eta_{l} }{\mu } \right)   \nonumber \\
			=&  \sum_{l=0}^{k} \Bigg[ (2-\gamma) \frac{(1-\phi)^{k-l} - (1-\gamma)^{k-l}}{\gamma - \phi} \nonumber \\
			&+ (1-\phi)^{k-l} \Bigg] \left( \frac{\alpha_{l} \delta}{\mu} + \frac{\alpha_{l} \eta_{l} }{\mu } \right) 
			\nonumber \\
			=&  \frac{1}{\gamma - \phi} \sum_{l=0}^{k} \Big[ (2-\phi)(1-\phi)^{k-l} \nonumber \\
			&  \quad - (2-\gamma)(1-\gamma)^{k-l}  \Big] \left( \frac{\alpha_{l} \delta}{\mu} + \frac{\alpha_{l} \eta_{l} }{\mu } \right) .
		\end{align}
		We first prove that $\eta_{k}$ is bounded, i.e., $\eta_{k} \leq D_{\eta}$. 
		We separate the sequence $\{\eta_{k}\}_{k\geq 0}$ into tow part. One is for $k < K$, and the other one is for $k \geq K$. For the first part, there always exists a bound for $\{ \eta_{k}\}_{0 \leq k < K}$ since it only has a finite number of elements. Therefore, we only need to prove the boundness of $\{ \eta_{k} \}_{k \geq K}$. We prove it by induction.
		Suppose that there exists an $k \geq K$ such that $\eta_{l} \leq D_{\eta}^{\prime}$ for all $K \leq l \leq k$, and it is sufficient to prove that $\eta_{k+1} \leq D_{\eta}^{\prime}$ by induction.
		
		We write~\eqref{eq: eta} as 
		\begin{align} \label{eq: seperate}
			\eta_{k+1} =& \frac{1}{\gamma - \phi} \sum_{l=0}^{K-1} \Big[ (2-\phi)(1-\phi)^{k-l} \nonumber \\
			& \quad - (2-\gamma)(1-\gamma)^{k-l}  \Big ] \left( \frac{\alpha_{l} \delta}{\mu} + \frac{\alpha_{l} \eta_{l} }{\mu } \right) \nonumber \\
			& + \frac{1}{\gamma - \phi} \sum_{l=K}^{k} \Big[ (2-\phi)(1-\phi)^{k-l} \nonumber \\
			& \quad - (2-\gamma)(1-\gamma)^{k-l} \Big ] \left( \frac{\alpha_{l} \delta}{\mu} + \frac{\alpha_{l} \eta_{l} }{\mu } \right).
		\end{align} 
		For the first term of~\eqref{eq: seperate}, there is 
		\begin{align*}
			& \sum_{l=0}^{K-1} \Big[ (2-\phi)(1-\phi)^{k-l} - (2-\gamma)(1-\gamma)^{k-l}  \Big ] \left( \frac{\alpha_{l} \delta}{\mu} + \frac{\alpha_{l} \eta_{l} }{\mu } \right) \\
			& \leq \frac{2(\gamma - \phi)}{\gamma \phi \mu} \max_{0\leq l < K}\{ \alpha_{l}\delta + \alpha_{l}\eta_{l} \}.
		\end{align*}
		Thus, we define $D_{\eta}^{''}:= \frac{2 \max_{0\leq l < K}\{ \alpha_{l}\delta + \alpha_{l}\eta_{l} \}}{\gamma \phi \mu}$. Since $\sum_{k=0}^{\infty} \alpha_{k} < \infty$, there exists a finite $\bar{\alpha}$ such that $\bar{\alpha} = \sup_{k} \alpha_{k}$. Define $D_{\eta}^{\prime} = \frac{\mu \gamma \phi D_{\eta}^{''} +  \bar{\alpha}\delta}{\mu \gamma \phi - \bar{\alpha}}$, then we have the following relationship based on~\eqref{eq: seperate}
		\begin{equation*}
			\eta_{k+1} \leq D_{\eta}^{''} + \frac{\bar{\alpha}(\delta + D_{\eta}^{\prime})}{\mu \gamma \phi} = D_{\eta}^{\prime}.
		\end{equation*} 
		Therefore, we can conclude that $\eta_{k} \leq D_{\eta}$ by letting $D_{\eta} = \inf_{K} \max \left\{ D_{\eta}^{\prime}, \max_{0 \leq l < K} \eta_{l}  \right\}$, $\forall k \geq 0$.

		Since $\eta_{k} \leq D_{\eta}$, we have the following result:
		\begin{align} \label{eq: sen1}
			\sum_{k=0}^{T} \frac{\| \Delta s_{i_{0},k} \|}{\theta_{\xi, k}} \leq & \sum_{k=0}^{T} \frac{\varphi_{k}}{\theta_{\xi, k}} \nonumber \\
			=& \sum_{k=1}^{T} \frac{1}{\theta_{\xi, k}} \sum_{l=0}^{k-1} (1-\gamma)^{k-1-l}\left( \frac{\alpha_{l} \delta}{\mu} + \frac{\alpha_{l} \eta_{l} }{\mu } \right) \nonumber \\
			\leq & \sum_{k=0}^{T-1} \frac{\alpha_{k} \delta + \alpha_{k} \eta_{k}}{\mu \theta_{\xi, k+1}} \sum_{l=0}^{T-1-k} (1-\gamma)^{l} \nonumber \\
			\leq & \frac{\delta + D_{\eta}}{\mu \gamma} \sum_{k=0}^{T-1}\frac{\alpha_{k}}{\theta_{\xi, k+1}} \nonumber \\
			\leq & \frac{\delta + D_{\eta}}{\mu \gamma} D_{\alpha,\xi} < \infty, \ \forall T \geq 0.
		\end{align}
		Similarly, we have
		\begin{align} \label{eq: sen2}
			&\sum_{k=0}^{T} \frac{\| \Delta \tilde{w}_{i_{0},k} \|}{\theta_{\zeta, k}} \leq  \sum_{k=0}^{T} \frac{\eta_{k}}{\theta_{\zeta, k}} \nonumber \\
			\leq & \sum_{k=1}^{T} \frac{\alpha_{k} \delta + \alpha_{k} \eta_{k}}{(\gamma - \phi) \mu \theta_{\zeta, k}} \sum_{l=0}^{T-1-k} [ (2-\phi)(1-\phi)^{k-l} \nonumber  \\
			&\quad - (2-\gamma)(1-\gamma)^{k-l}  ] \nonumber \\
			\leq & \frac{\delta + D_{\eta}}{(\gamma - \phi)\mu}\left(\frac{2-\phi}{\phi} - \frac{2-\gamma}{\gamma} \right) \sum_{k=0}^{T-1}\frac{\alpha_{k}}{\theta_{\zeta, k+1}} \nonumber \\
			\leq & \frac{\delta + D_{\eta}}{\mu \gamma \phi} D_{\alpha, \zeta} < \infty, \ \forall T \geq 0.
		\end{align}
		
		
		From Algorithm~\ref{algo: one}, recall that we fixed the observation sequence, the probability comes from the noise $\boldsymbol{\xi}_{k}$ and $\boldsymbol{\zeta}_{k}$. 
		Therefore, the probability of execution is reduced to 
		\begin{equation*}
			\mathbb{P}[\mathcal{R}^{-1}(\mathcal{P}, \mathcal{O}, \mathbf{s}_{0}, \tilde{\mathbf{w}}_{0}, \mathbf{w}_{0})] = \iint f_{\xi \zeta}(\boldsymbol{\xi}, \boldsymbol{\zeta}) d\boldsymbol{\xi} d \boldsymbol{\zeta},
		\end{equation*}
		where $\iint f_{\xi \zeta}(\boldsymbol{\xi}, \boldsymbol{\zeta}) d\boldsymbol{\xi} d \boldsymbol{\zeta} = \prod \limits_{i=1}^{N} \prod \limits_{k=0}^{\infty} f_{L}(\xi_{i,k}, \theta_{\xi_{i},k}) f_{L}(\zeta_{i,k}, \theta_{\zeta_{i},k})$. According to~\eqref{eq: equal_noise}, \eqref{eq: noise}, \eqref{eq: sen1} and~\eqref{eq: sen2}, we derive 
		\begin{align*}
			&\frac{\mathbb{P}[\mathcal{R}^{-1}(\mathcal{P}, \mathcal{O}, \mathbf{s}_{0}, \tilde{\mathbf{w}}_{0}, \mathbf{w}_{0})]}{\mathbb{P}[\mathcal{R}^{-1}(\mathcal{P}^{\prime}, \mathcal{O}, \mathbf{s}_{0}, \tilde{\mathbf{w}}_{0}, \mathbf{w}_{0})]} \\
			=& \prod \limits_{k=0}^{\infty} \frac{f_{L}(\xi_{i_{0},k}, \theta_{\xi_{i_{0},k}}) f_{L}(\zeta_{i_{0},k}, \theta_{\zeta_{i_{0},k}})}{f_{L}(\xi_{i_{0},k}^{\prime}, \theta_{\xi_{i_{0},k}}) f_{L}(\zeta_{i_{0},k}^{\prime}, \theta_{\zeta_{i_{0},k}})} \\
			\leq & \prod \limits_{k=0}^{\infty} e^{ \frac{\| \Delta \xi_{i_{0},k} \|_{1}}{\theta_{\xi,k}} + \frac{\| \Delta \zeta_{i_{0},k} \|_{1}}{\theta_{\zeta,k}} } \\
			=& \exp  \sum_{k=0}^{\infty} \left( \frac{\| \Delta s_{i_{0},k} \|_{1}}{\theta_{\xi,k}} + \frac{\| \Delta \tilde{w}_{i_{0},k} \|_{1}}{\theta_{\zeta,k}}   \right) \\
			\leq & \exp \left( \frac{\delta + D_{\eta}}{\mu \gamma \phi} (D_{\alpha, \xi} + \phi D_{\alpha, \zeta})  \right).
		\end{align*} 
		Therefore, we have $\epsilon = \frac{\delta + D_{\eta}}{\mu \gamma \phi} (D_{\alpha, \xi} + \phi D_{\alpha, \zeta})$.
	\end{pf}
	
	From Theorem~\ref{thm: DP}, we observe that the privacy level $\epsilon$ is proportional to $D_{\alpha, \xi} = \sum_{k=0}^{\infty} \frac{\alpha_{k}}{\theta_{\xi,k}}$ and $D_{\alpha, \zeta} = \sum_{k=0}^{\infty} \frac{\alpha_{k}}{\theta_{\zeta,k}}$. 
Note that $\alpha_{k}$ reflects the mismatch between supply and demand. To enhance the privacy of DP-DGT, we can reduce step sizes or increase noise power, but this comes at the cost of accuracy. Therefore, there is a trade-off between privacy and convergence accuracy.
	
	We summarize the theoretical results in Theorem~\ref{thm: primal_con} and~\ref{thm: DP} and conclude that it is possible to choose the parameters $\alpha_{k}$, $\theta_{\xi,k}$ and $\theta_{\zeta,k}$ such that DP-DGT can lead $\{\mathbf{w}_{k}\}$ to converge to a neighborhood of $\mathbf{w}^{*}$ almost surely while achieving $\epsilon$-differential privacy.
	\begin{corollary} \label{cor: combine}
		Consider DP-DGT under Assumptions~\ref{assum: convexity_slater}--\ref{assum: noise}. When		$\sum_{k=0}^{\infty} \alpha_{k} < \infty$, 
		$\sum_{k=0}^{\infty}\theta_{\xi,k}^{2} < \infty$, 
		$\sum_{k=0}^{\infty}\theta_{\zeta,k}^{2} < \infty$, 
		$\sum_{k=0}^{\infty} \frac{\alpha_{k}}{\theta_{\xi,k}} < \infty$, 
		$\sum_{k=0}^{\infty} \frac{\alpha_{k}}{\theta_{\zeta,k}} < \infty$, 
		$\sum_{k=0}^{\infty} \frac{\theta_{\xi,k}^{2}}{\alpha_{k}}< \infty$ 
		$\sum_{k=0}^{\infty} \frac{\theta_{\zeta,k}^{2}}{\alpha_{k}}< \infty$,
		and there exists $\lambda$ satisfying $q_{C} < \lambda< 1$ and $q_{R} < \lambda< 1$ and $k_{0}>0$ such that $\frac{\alpha_{k}}{\alpha_{k_0}} \geq \beta \lambda^{k-k_{0}}$, then $\{\mathbf{w}_{k}\}$ converges to a neighborhood of $\mathbf{w}^{*}$ almost surely and while achieving $\epsilon$-differential privacy simultaneously.
	\end{corollary}
	There indeed exists a possibility of choosing $\alpha_{k}$, $\theta_{\xi,k}$ and $\theta_{\zeta,k}$ such that all conditions listed in Corollary~\ref{cor: combine} can be satisfied simultaneously. For example, we can let $\alpha_{k}$, $\theta_{\xi,k}$ and $\theta_{\zeta,k}$ decrease linearly and further derive a close form of expression of $\epsilon$.
	
	\begin{corollary} \label{cor: finite}
		Consider DP-DGT under Assumptions~\ref{assum: convexity_slater}--\ref{assum: noise}. Let $\alpha_{k} = \alpha_{0}q^{k}$, $\theta_{\xi,k} = \theta_{\xi,0}q_{\xi}^{k}$, and $\theta_{\zeta,k} = \theta_{\zeta,0}q_{\zeta}^{k}$. 
		If $\alpha_{0} < \mu \gamma \phi$, and $\{q_{R}, q_{C}, q_{\xi}^{2}, q_{\zeta}^{2} \} < q < \{ q_{\xi}, q_{\zeta} \} < 1$, then 
		\begin{equation} \label{eq: spec}
			\epsilon = \frac{\alpha_{0}\delta(\gamma \phi \mu + \alpha_{0})}{\gamma \phi \mu (\gamma \phi \mu - \alpha_{0})}\left( \frac{q_{\xi}}{\theta_{\xi,0}(q_{\xi} - q)} + \frac{\phi q_{\zeta}}{\theta_{\zeta,0}(q_{\zeta} - q)}  \right),
		\end{equation}
        with the error bound
        \begin{equation} \label{eq: error_bound} 
            \begin{aligned}
            & \mathbb{E}[ \| \mathbf{w}_{k} - \mathbf{w}^{*}  \|^{2}] \\
    \leq &  O \left( 1 + \frac{1}{(q - q_{\zeta}^{2})(q - q_{\xi}^{2})(q-q_{c})(q-q_{R})(1-q)}   \right).
    \end{aligned}
        \end{equation}
	\end{corollary}
\begin{pf}
		The proof is provided in Appendix A.4~(\cite{huo2024differentially}).
\end{pf}

As shown in previous works using $\delta$-adjacency~(\cite{ding2021differentially, ding2021differentiallyra}), the privacy loss is proportional to $\delta$ in~\eqref{eq: spec}. The value of $\delta$ reflects the distance between adjacent problems, indicating that larger differences require more noise for privacy preservation. Additionally, the result in~\eqref{eq: error_bound} highlights the trade-off between noise level and convergence accuracy.
	\begin{remark}
In contrast to \cite{chen2023differentially}, where cumulative privacy loss increases indefinitely, the cumulative privacy loss $\epsilon$ in~\eqref{eq: spec} remains constant even as the number of iterations grows. Additionally, unlike \cite{chen2023differentially} and \cite{huang2015differentially}, we derive this result without assuming bounded gradients.
	\end{remark}
	
	\section{Numerical Simulations} \label{sec: sim}
The future microgrid is evolving into a cyber-physical system with three layers (\cite{huang2010future}): a physical layer (power network), a communication layer (information transmission network), and a control layer (running distributed algorithms and processing communication data). Each bus node has a corresponding agent, and they exchange information via the communication network. We consider an economic dispatch problem for the IEEE 14-bus power system, where the generator buses are $\{1, 2, 3, 6, 8 \}$ and the load buses are $\{2, 3, 4, 5, 6, 9, 10, 11, 13, 14 \}$. Notably, the communication network among buses can be independent of the actual bus connections (\cite{yang2013consensus}). 
	We model the directed communication network as $\mathcal{G} = (\mathcal{V}, \mathcal{E})$, where $\mathcal{V}$ is the set combing the generators buses and load buses, and $\mathcal{E} = \{ (i,i+1), (i,i+2)| 1 \leq i \leq 12 \} \cup \{ (13,14), (13,1), (14,1), (1,7), (2,8), (3,2), (3,9), (4,10), \\ (5,2), (5,11), (6,12) \}$. 
	The cost functions of the generator $i$ is
		$F_{i}(w_{i}) = a_{i}w_{i}^{2} + b_{i}w_{i} + c_{i}$,
	The generator parameters, including the parameters of the quadratic cost functions, are adapted from \cite{kar2012distributed} and presented in Table~\ref{tab: IEEE_14}. 
	\begin{table}[t] 
		\caption{IEEE 14-bus system generator parameters}
		\label{tab: IEEE_14}
		\centering
		\tiny
		\begin{tabular}{m{0.05\textwidth}<{\centering}m{0.07\textwidth}<{\centering}m{0.07\textwidth}<{\centering}m{0.07\textwidth}<{\centering}}
			\toprule
			Bus & $a_{i}$(MW$^{2}$h) & $b_{i}$(\$/MWh) & Range(MW) \\
			\midrule
			$1$ & $0.04$ & $2.0$ & $[0,80]$ \\
			$2$ & $0.03$ & $3.0$ & $[0,90]$ \\
			$3$ & $0.035$ & $4.0$ & $[0,70]$ \\
			$6$ & $0.03$ & $4.0$ & $[0,70]$ \\
			$8$ & $0.04$ & $2.5$ & $[0,80]$ \\
			\bottomrule
		\end{tabular}
	\end{table}
	When a bus does not contain generators, the power generation at that bus is set to zero. Thus, the update in~\eqref{eq: primal_update} simply becomes $w_{i,k} = 0$ for $i \notin \{1, 2, 3, 6, 8 \}$.
	The virtual local demands at each bus are given as $D_{1} = 0\ \text{MW}$, $D_{2} = 9\ \text{MW}$, $D_{3} = 56\ \text{MW}$, $D_{4} = 55\ \text{MW}$, $D_{5} = 27\ \text{MW}$, $D_{6} = 27\ \text{MW}$, $D_{7} = 0\ \text{MW}$, $D_{8} = 0\ \text{MW}$, $D_{9} = 8\ \text{MW}$, $D_{10} = 24\ \text{MW}$, $D_{11} = 53\ \text{MW}$, $D_{12} = 46\ \text{MW}$, $D_{13} = 16\ \text{MW}$, and $D_{14} = 40\ \text{MW}$. The total demand is $D = \sum_{i=1}^{14} D_{i} = 361 \ \text{MW}$, which is unknown to the agent at each bus. The optimal solution $\mathbf{w}^{*}$ is obtained by using the CVX solver in a centralized manner, which is $[76.7398, 85.6530, 59.1311, 68.9863, 70.4898]^{T}$.

	\subsection{Convergence of DP-DGT}
\begin{figure}[t]  
		\centering
		\includegraphics[width=0.8\linewidth]{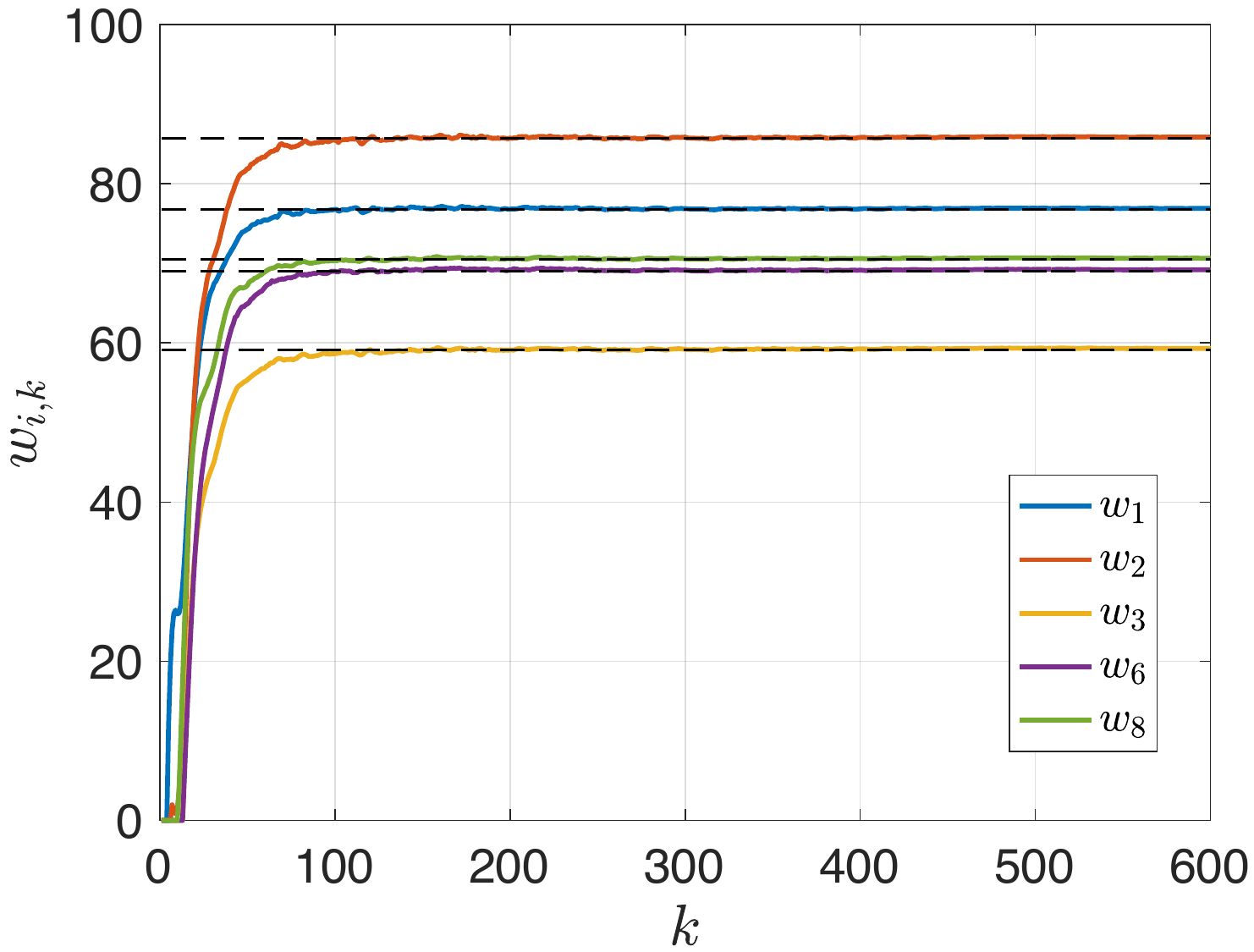}
		\caption{Convergence performance of the generated power from Generators 1, 2, 3, 6, and 8.} 
		\label{fig: expect_con}
	\end{figure}

\begin{figure}[t]  
		\centering
		\includegraphics[width=0.8\linewidth]{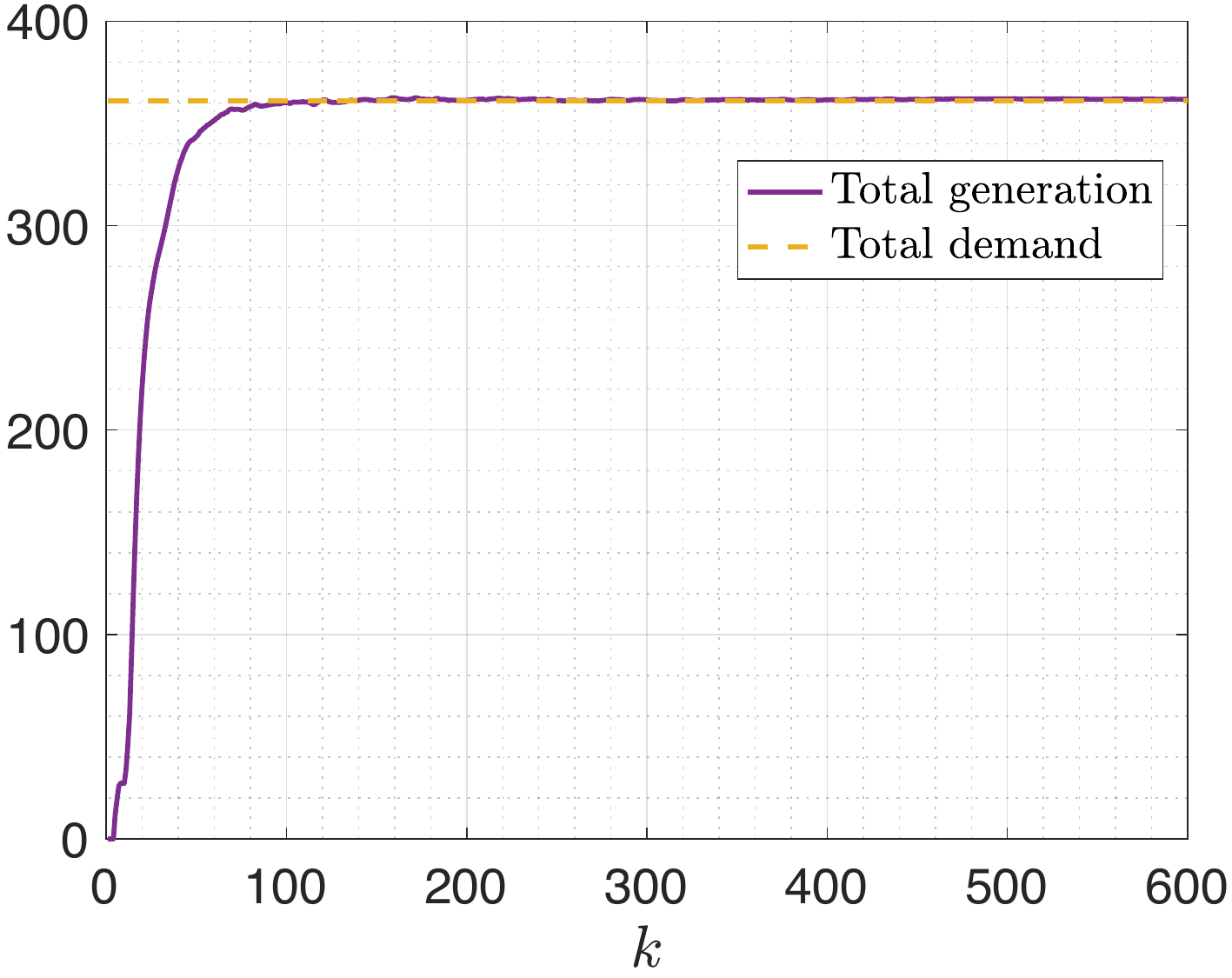}
		\caption{Total generated power and demand.} 
		\label{fig: s_d}
	\end{figure}

	We set the step size parameters to $\alpha_{0} = 0.015$ and $q = 0.991$. 
	The Laplacian noise are chosen as $\theta_{\xi,0} = \theta_{\zeta,0} = 0.01$ and $q_{\xi} = q_{\zeta} = 0.995$. 
	Additionally, we select $\gamma = 0.8$ and $\phi = 0.7$. 
	Fig.~\ref{fig: expect_con} shows that the decisions of generator buses $1$, $2$, $3$, $6$, and $8$ converge to small neighborhoods of the optimal allocation solutions.
	Fig.~\ref{fig: s_d} shows that the total generation asymptomatically converges to a neighborhood of the total demand. 

	\subsection{Tradeoff between Accuracy and Privacy}
\begin{figure}[t]  
		\centering
		\includegraphics[width=0.8\linewidth]{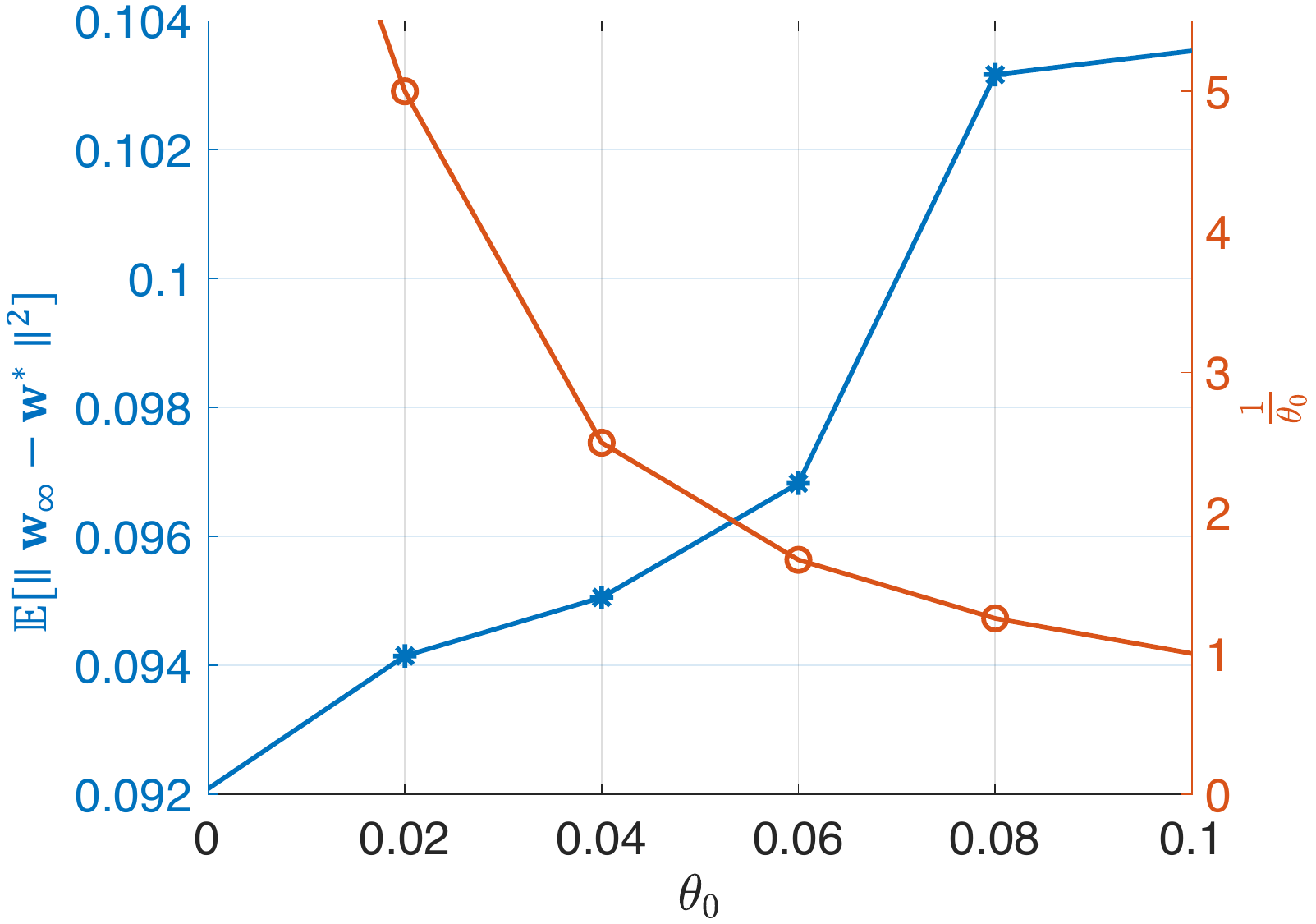}
		\caption{Variations of $\mathbb{E}\left[\|\mathbf{w}_{\infty} - \mathbf{w}^{*} \|^{2}\right]$ and the trend of $\epsilon$ with different noise intensity $\theta_{0}$.} 
		\label{fig: tradeoff}
	\end{figure}

	To demonstrate the tradeoff between convergence accuracy and the privacy level. We let $\theta_{\xi,0} = \theta_{\zeta,0} = \theta_{0}$, fix $\alpha_{0} = 0.015$, $q = 0.991$, $q_{\xi} = q_{\zeta} = 0.995$, and vary $\theta_{0}$ from $0$ to $0.1$. 
	Due to the randomness of the Laplacian noise, we run the simulation $2000$ times and obtain the empirical mean.
	Fig.~\ref{fig: tradeoff} plots $\mathbb{E}\left[\|\mathbf{w}_{\infty} - \mathbf{w}^{*} \|^{2}\right]$ and $\frac{1}{\theta_{0}}$ under different intense of Laplacian noise, where the latter represents the trend of $\epsilon$. 
	Roughly speaking, Fig.~\ref{fig: tradeoff} shows that as $\theta_{0}$ increases, the expected convergence error $\mathbb{E}\left[\| \mathbf{w}_{\infty} - \mathbf{w}^{*} \|^{2}\right]$ increases. Moreover, as $\theta_{0}$ increases, $\epsilon$ increases as well. Thus, the tradeoff between the privacy level and the convergence is illustrated.
	
	\subsection{Comparison with State-of-the-art}
	We further compare the proposed algorithm with the conventional distributed dual gradient tracking algorithm (DDGT) expressed as~\eqref{eq: alg_conv} in \cite{zhang2020distributed} and the differentially private distributed resource allocation via deviation tracking algorithm (diff-DRADT) in \cite{ding2021differentiallyra}. 
	The step size and the noise is set to $\alpha_{k} = 0.034 \times 0.99^{k}$ and $\theta_{\xi,k} = \theta_{\zeta,k} = 0.01 \times 0.995^{k}$, respectively.
	Since there is no privacy preservation in conventional DDGT, to be fair, we run it with the same noise parameter and let $\iota \beta_{k} = \alpha_{k} = 0.034 \times 0.99^{k}$.
	Specifically, we set $\beta_{k} = 0.99^{k}$ and $\iota = 0.034$.
	\cite{ding2021differentiallyra} used the constant step size and linearly decreasing noise to achieve the finite cumulative privacy loss. Hence, we set the noise the same as ours and set the step size as 0.015.  
	\begin{figure}[t]  
		\centering
		\includegraphics[width=0.8\linewidth]{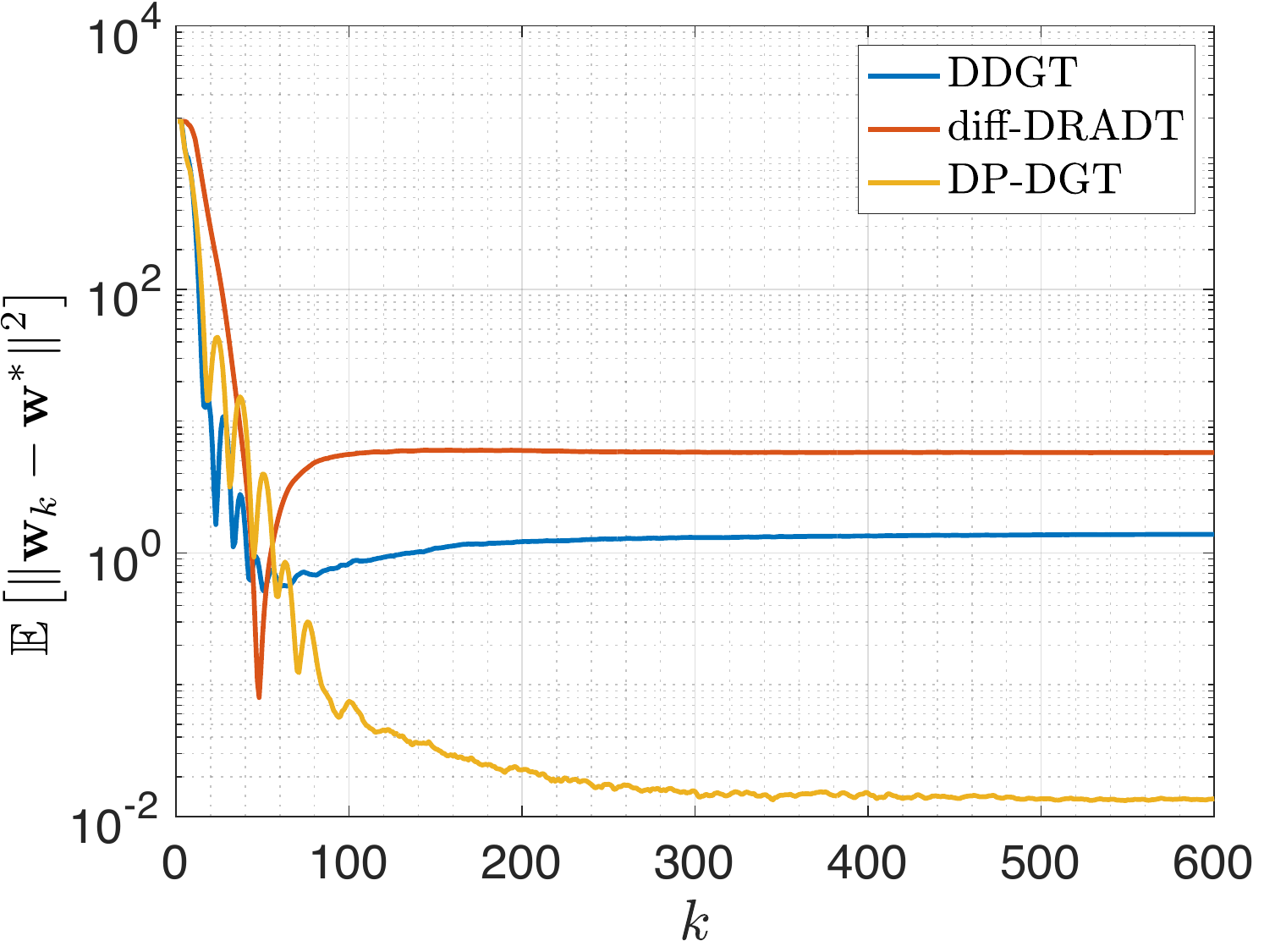}
		\caption{Comparison of our proposed algorithm with the conventional DDGT algorithm~(\cite{zhang2020distributed}) with the same privacy level and the diff-DRADT in \cite{ding2021differentiallyra}.} 
		\label{fig: compare}
	\end{figure}
	Fig.~\ref{fig: compare} depicts the comparison results. Since diff-DRADT only considers undirected graphs, it suffers from a high optimization error in the directed graph. Due to the lack of robustness, directly adding noise in DDGT will cause noise accumulation and accuracy compromise. It can be seen that under the same noise, our algorithm achieves the best convergence accuracy.

	\section{Conclusion and Future Work} \label{sec: conclusions}
	This paper investigates privacy preservation in distributed resource allocation problems over directed unbalanced networks. 
	We propose a novel differentially private distributed deviation tracking algorithm that incorporates noises into transmitted messages to ensure privacy. 
	The distribution of noise and step sizes are carefully designed to guarantee convergence and achieve $\epsilon$-differential privacy simultaneously. 
	
	Potential future research directions include exploring methods to relax the step size requirement and accelerate convergence. 
	Another interesting topic is to deal with the tradeoff between convergence accuracy and privacy level.
	
\bibliographystyle{agsm}
\bibliography{HW}  

	 \appendix

    	\section{Appendix}
	\subsection{Preliminary Lemma}
	\begin{lem} \label{lem: pre_con}
		~\cite{robbins1971convergence}
		Let $\{ u_{k}\}$, $\{ v_{k}\}$, $\{ w_{k} \}$ and $\{ z_{k} \}$ be the nonnegative sequences of random variables. If they satisfy
		\begin{align*}
			\mathbb{E}[ u_{k+1} ] \leq  (1+ z_{k})u_{k} - v_{k} + w_{k}, \\
			\sum_{k=0}^{\infty} z_{k} < \infty \ a.s., \ \text{and} \  \sum_{k=0}^{\infty} w_{k} < \infty \ a.s.,
		\end{align*}
		then $u_{k}$ converges almost surely (a.s.) to a finite value, and $\sum_{k=0}^{\infty} v_{k} < \infty$ a.s..
	\end{lem}
	\subsection{Proof of Lemma~\ref{lem: LMI}} \label{app: LMI}
	
	We aim to prove the component-wise inequalities in~\eqref{eq: LMI}.  
	To simplify the presentation, we replace the notations $I_{N}$ and $\mathbf{1}_{N}$ as $I$ and $\mathbf{\mathbf{1}}$, respectively.
	
	\noindent \textbf{\romannumeral1) Bound $\mathbb{E}\left[\left\| \mathbf{x}_{k} - \mathbf{1} \bar{\mathbf{x}}_{k}^{T}  \right\|^{2}_{R}  \big| \mathcal{F}_{k} \right]$ and obtain the first inequality:}
	
	From~\eqref{eq: x_compact}, we can derive that
	\begin{equation*}
		\begin{aligned}
			& \mathbf{x}_{k+1} - \mathbf{1} \bar{\mathbf{x}}_{k+1}^{T} \\
			=& (\mathbf{R}_{\phi} - \mathbf{1}\pi_{\mathbf{R}}^{T})(\mathbf{x}_{k} - \mathbf{1} \bar{\mathbf{x}}_{k}^{T}) -  (I - \mathbf{1}\pi_{\mathbf{R}}^{T}) \mathbf{v}_{k} \\ 
			&- \phi (I - \mathbf{1}\pi_{\mathbf{R}}^{T}) \mathbf{R}\boldsymbol{\zeta}_{k} -  \gamma (I - \mathbf{1}\pi_{\mathbf{R}}^{T}) \mathbf{C}\boldsymbol{\xi}_{k}.
		\end{aligned}
	\end{equation*}
	Therefore, we have
	\begin{align} \label{eq: x_1}
		&\mathbb{E}\left[\left\| \mathbf{x}_{k+1} - \mathbf{1} \bar{\mathbf{x}}_{k+1}^{T}  \right\|^{2}_{R} \big| \mathcal{F}_{k} \right] \nonumber \\
		= & \sigma_{R}^{2} \left\| \mathbf{x}_{k} - \mathbf{1} \bar{\mathbf{x}}_{k}^{T} \right\|^{2}_{R} +  \left\| I - \mathbf{1}\pi_{\mathbf{R}}^{T} \right\|_{R}^{2} \mathbb{E}\left[\left\| \mathbf{v}_{k} \right\|_{R}^{2} \big| \mathcal{F}_{k} \right] \nonumber \\
		& + 2 \sigma_{R}\left\| I - \mathbf{1}\pi_{\mathbf{R}}^{T} \right\|_{R} \left\| \mathbf{x}_{k} - \mathbf{1} \bar{\mathbf{x}}_{k}^{T} \right\|_{R} \left\| \mathbf{v}_{k} \right\|_{R} \nonumber \\
		&+ \phi^{2} \left\|\mathbf{R} - \mathbf{1}\pi_{\mathbf{R}}^{T} \right\|^{2}_{R} \mathbb{E}\left[ \left\| \boldsymbol{\zeta}_{k} \right\|^{2}_{R} \right] \nonumber \\
		&+ \gamma^{2}  \left\|(I - \mathbf{1}\pi_{\mathbf{R}}^{T}) \mathbf{C} \right\|^{2}_{R}\mathbb{E}[\left\| \boldsymbol{\xi}_{k} \right\|^{2}_{R}] 
		\nonumber \\
		\leq & \frac{1+\sigma_{R}^{2}}{2} \left\| \mathbf{x}_{k} - \mathbf{1} \bar{\mathbf{x}}_{k}^{T}  \right\|^{2}_{R} \nonumber \\
		&+  \frac{1+\sigma_{R}^{2}}{1-\sigma_{R}^{2}} \left\| I - \mathbf{1}\pi_{\mathbf{R}}^{T} \right\|_{R}^{2} \mathbb{E}\left[\left\| \mathbf{v}_{k} \right\|_{R}^{2} \big| \mathcal{F}_{k} \right] \nonumber \\
		& + \phi^{2} Nm \left\|\mathbf{R} - \mathbf{1}\pi_{\mathbf{R}}^{T} \right\|^{2}_{R} \theta_{\zeta,k}^{2} \nonumber \\
		& + \gamma^{2} Nm  \left\|(I - \mathbf{1}\pi_{\mathbf{R}}^{T}) \mathbf{C} \right\|^{2}_{R} \theta_{\xi,k}^{2},
	\end{align}
	where the equality is based on the dependence of the added noise and Lemma~\ref{lem: sigma}.
	For the second term, we have
	\begin{align} \label{eq: x_2}
		\left\| \mathbf{v}_{k} \right\|_{R}^{2} =& \left\| \mathbf{v}_{k} - \pi_{\mathbf{C}}\hat{\mathbf{v}}_{k}^{T} + \pi_{\mathbf{C}}\hat{\mathbf{v}}_{k}^{T}\right\|_{R}^{2} \nonumber \\
		\leq & 2 \delta_{R, C} \left\| \mathbf{v}_{k} - \pi_{\mathbf{C}}\hat{\mathbf{v}}_{k}^{T}\right\|_{C}^{2} + 2 \left\| \pi_{\mathbf{C}} \right\|_{R}^{2} \left\| \hat{\mathbf{v}}_{k}^{T}\right\|_{R}^{2},
	\end{align}
	and   
	\begin{align} \label{eq: x_3}
		&\left\| \hat{\mathbf{v}}_{k}^{T}\right\|_{R}^{2} \nonumber \\
		=& \left\| \alpha_{k} \mathbf{1}^{T} G(\mathbf{x}_{k}) - \alpha_{k} \mathbf{1}^{T} G( \mathbf{1} \bar{\mathbf{x}}_{k}^{T}) + \alpha_{k}  \mathbf{1}^{T} G( \mathbf{1} \bar{\mathbf{x}}_{k}^{T})  \right\|_{R}^{2}  \nonumber \\
		\leq & 2L^{2}N \alpha_{k}^{2} \left\| \mathbf{x}_{k} - \mathbf{1} \bar{\mathbf{x}}_{k}^{T} \right\|_{R}^{2} + 2 \alpha_{k}^{2} \left\| \nabla f(\bar{\mathbf{x}}_{k})\right\|_{2}^{2},
	\end{align}
	where the inequality is based on the $L$-Lipschitz smoothness of $f_{i}$.                                                     
	Combining~\eqref{eq: x_1}--\eqref{eq: x_3}, we obtain
	\begin{align}
		&\mathbb{E}\left[\left\| \mathbf{x}_{k+1} - \mathbf{1} \bar{\mathbf{x}}_{k+1}^{T}  \right\|^{2}_{R} \big| \mathcal{F}_{k} \right] \nonumber \\
		\leq & \left( \frac{1+\sigma_{R}^{2}}{2} + \mathbf{a}_{1} \alpha_{k}^{2} \right) \left\| \mathbf{x}_{k} - \mathbf{1} \bar{\mathbf{x}}_{k}^{T}  \right\|^{2}_{R} \nonumber \\
		& + 2\frac{1+\sigma_{R}^{2}}{1-\sigma_{R}^{2}} \left\| I_{N} - \mathbf{1}\pi_{\mathbf{R}}^{T} \right\|_{R}^{2} \delta_{R,C}^{2}   \left\| \mathbf{v}_{k} - \pi_{\mathbf{C}}\hat{\mathbf{v}}_{k}^{T}\right\|_{\mathbf{C}}^{2}  \nonumber \\
		& + 4 \frac{1+\sigma_{R}^{2}}{1-\sigma_{R}^{2}} \left\| I_{N} - \mathbf{1}\pi_{\mathbf{R}}^{T} \right\|_{R}^{2}  
		\left\| \pi_{\mathbf{C}} \right\|_{R}^{2} \alpha_{k}^{2}
		\mathbb{E}[\left\| \nabla f(\bar{\mathbf{x}}_{k})\right\|_{2}^{2}] \nonumber \\
		& + \phi^{2} Nm \left\| \mathbf{R} - \mathbf{1}\pi_{\mathbf{R}}^{T} \right\|_{R}^{2}  \theta_{\zeta,k}^{2} \nonumber \\
		& + \gamma^{2} Nm  \left\| (I - \mathbf{1}\pi_{\mathbf{R}}^{T}) \mathbf{C} \right\|_{R}^{2}  \theta_{\xi,k}^{2}.
	\end{align}

	\noindent \textbf{\romannumeral2) Bound $\mathbb{E}\left[\left\| \mathbf{v}_{k} - \pi_{\mathbf{C}}\hat{\mathbf{v}}_{k}^{T} \right\|_{C}^{2} \big | \mathcal{F}_{k} \right]$ and get the second inequality:}
	
	From~\eqref{eq: y_compact}, we can obtain that
	\begin{equation*} 
		\begin{aligned}
			&\mathbf{v}_{k+1} - \pi_{\mathbf{C}}\hat{\mathbf{v}}_{k+1}^{T} \\
			=& (\mathbf{C}_{\gamma} - \pi_{\mathbf{C}}\mathbf{1}^{T})(\mathbf{v}_{k} - \pi_{\mathbf{C}}\hat{\mathbf{v}}_{k}^{T}) + (I - \pi_{\mathbf{C}})\mathbf{v}_{k+1}  \\
			&- (\mathbf{C}_{\gamma} - \pi_{\mathbf{C}}\mathbf{1}^{T}) \mathbf{v}_{k} \\
			= & (\mathbf{C}_{\gamma} - \pi_{\mathbf{C}}\mathbf{1}^{T})(\mathbf{v}_{k} - \pi_{\mathbf{C}}\hat{\mathbf{v}}_{k}^{T}) - (\mathbf{C}_{\gamma} - \pi_{\mathbf{C}}\mathbf{1}^{T}) \mathbf{v}_{k}   \\
			& + (I - \pi_{\mathbf{C}} \mathbf{1}^{T} )[ (\mathbf{C}_{\gamma} - I) \mathbf{y}_{k+1} + \alpha_{k+1}G(\mathbf{x}_{k+1}) ]   \\
			=&  (\mathbf{C}_{\gamma} - \pi_{\mathbf{C}}\mathbf{1}^{T})(\mathbf{v}_{k} - \pi_{\mathbf{C}}\hat{\mathbf{v}}_{k}^{T}) + (I - \pi_{\mathbf{C}} \mathbf{1}^{T} ) \alpha_{k+1}G(\mathbf{x}_{k+1}) \\
			& - (\mathbf{C}_{\gamma} - \pi_{\mathbf{C}}\mathbf{1}^{T})[(\mathbf{C}_{\gamma}-I) \mathbf{y}_{k} + \alpha_{k}G(\mathbf{x}_{k}) ]  \\
			& + (I - \pi_{\mathbf{C}} \mathbf{1}^{T} ) (\mathbf{C}_{\gamma} - I) (\mathbf{C}_{\gamma}\mathbf{y}_{k} + \alpha_{k}G(\mathbf{x}_{k}) +\gamma\mathbf{C}\boldsymbol{\xi}_{k} )  \\
			=& (\mathbf{C}_{\gamma} - \pi_{\mathbf{C}}\mathbf{1}^{T})(\mathbf{v}_{k} - \pi_{\mathbf{C}}\hat{\mathbf{v}}_{k}^{T}) \\
			&+ (I - \pi_{\mathbf{C}}\mathbf{1}^{T})( \alpha_{k+1} G(\mathbf{x}_{k+1}) - \alpha_{k} G(\mathbf{x}_{k})) + \gamma (\mathbf{C}_{\gamma} - I)\mathbf{C}\boldsymbol{\xi}_{k}.
		\end{aligned}	
	\end{equation*}
	Therefore, we have
	\begin{align} \label{eq: v_1}
		& \mathbb{E}\left[\left\| \mathbf{v}_{k+1} - \pi_{\mathbf{C}}\hat{\mathbf{v}}_{k+1}^{T} \right\|_{C}^{2} \big |\mathcal{F}_{k} \right] \nonumber 	\\
		\leq & \frac{1+\sigma_{C}^{2}}{2}\left\| \mathbf{v}_{k} - \pi_{\mathbf{C}}\hat{\mathbf{v}}_{k}^{T} \right\|_{C}^{2} \nonumber \\
		&+ \frac{1+\sigma_{C}^{2}}{1-\sigma_{C}^{2}} \left\| I_{N} - \pi_{\mathbf{C}}\mathbf{1}^{T}\right\|_{C}^{2}
		\mathbb{E}\big[ \left\| \alpha_{k+1} G(\mathbf{x}_{k+1}) - \alpha_{k} G(\mathbf{x}_{k}) \right. \nonumber \\
		&+ \left. \gamma (\mathbf{C}_{\gamma} - I)\mathbf{C}\boldsymbol{\xi}_{k} \right\|_{C}^{2} \big] \nonumber \\
		\leq & \frac{1+\sigma_{C}^{2}}{2}\left\| \mathbf{v}_{k} - \pi_{\mathbf{C}}\hat{\mathbf{v}}_{k}^{T} \right\|_{C}^{2} \nonumber \\
		&+ \frac{1+\sigma_{C}^{2}}{1-\sigma_{C}^{2}} \left\| I_{N} - \pi_{\mathbf{C}}\mathbf{1}^{T}\right\|_{C}^{2} \mathbb{E}\left[ \left\| \alpha_{k+1} G(\mathbf{x}_{k+1}) - \alpha_{k} G(\mathbf{x}_{k}) \right\|_{C}^{2} \right] \nonumber \\
		&+ 2 \frac{1+\sigma_{C}^{2}}{1-\sigma_{C}^{2}} \left\| I_{N} - \pi_{\mathbf{C}}\mathbf{1}^{T}\right\|_{C}^{2} \mathbb{E}\left[\| \gamma ( \mathbf{C}_{\gamma} - I)C \boldsymbol{\xi}_{k}\|_{C}^{2}\right].
	\end{align}
	For the third term of~\eqref{eq: v_1}, we have
	\begin{align} \label{eq: v_2}
		&\mathbb{E}[\gamma (\alpha_{k+1}G(\mathbf{x}_{k+1}) - \alpha_{k} G(\mathbf{x}_{k}))^{T}(\mathbf{C}_{\gamma} - I)C \boldsymbol{\xi}_{k}| \mathcal{F}_{k}] \nonumber \\
		=& \mathbb{E}[\gamma \alpha_{k+1}G(\mathbf{x}_{k+1})^{T}(\mathbf{C}_{\gamma} - I)C \boldsymbol{\xi}_{k}| \mathcal{F}_{k}] \nonumber\\
		=& \mathbb{E}[\gamma \alpha_{k+1}G(\mathbf{R}_{\phi} \mathbf{x}_{k} -  \mathbf{v}_{k} - \phi \mathbf{R} \boldsymbol{\zeta}_{k} - \gamma  \mathbf{C} \boldsymbol{\xi}_{k})^{T}(\mathbf{C}_{\gamma} - I)C \boldsymbol{\xi}_{k}| \mathcal{F}_{k}] \nonumber\\
		\leq & 2 \gamma^{2}LNm \alpha_{k+1} \|\mathbf{C} \| \| (\mathbf{C}_{\gamma}-I)\mathbf{C} \| \theta_{\xi,k}^{2},
	\end{align}
	where the inequality is from $\mathbb{E}[\gamma \alpha_{k+1}G(\mathbf{R}_{\phi} \mathbf{x}_{k} -  \mathbf{v}_{k} - \phi \mathbf{R} \boldsymbol{\zeta}_{k} - \gamma  \mathbf{C} \boldsymbol{\xi}_{k})^{T}(\mathbf{C}_{\gamma} - I)C \boldsymbol{\xi}_{k}| \mathcal{F}_{k}] = \gamma \alpha_{k+1} \mathbb{E}[( G(\mathbf{R}_{\phi} \mathbf{x}_{k} -  \mathbf{v}_{k} - \phi \mathbf{R} \boldsymbol{\zeta}_{k} - \gamma  \mathbf{C} \boldsymbol{\xi}_{k}) - G(\mathbf{R}_{\phi} \mathbf{x}_{k} -  \mathbf{v}_{k} - \phi \mathbf{R} \boldsymbol{\zeta}_{k}) + G(\mathbf{R}_{\phi} \mathbf{x}_{k} -  \mathbf{v}_{k} - \phi \mathbf{R} \boldsymbol{\zeta}_{k}) )^{T} (\mathbf{C}_{\gamma} - I)C \boldsymbol{\xi}_{k} | \mathcal{F}_{k} ] = \mathbb{E}[( G(\mathbf{R}_{\phi} \mathbf{x}_{k} -  \mathbf{v}_{k} - \phi \mathbf{R} \boldsymbol{\zeta}_{k} - \gamma  \mathbf{C} \boldsymbol{\xi}_{k}) - G(\mathbf{R}_{\phi} \mathbf{x}_{k} -  \mathbf{v}_{k} - \phi \mathbf{R} \boldsymbol{\zeta}_{k}) )^{T} (\mathbf{C}_{\gamma} - I)C \boldsymbol{\xi}_{k} | \mathcal{F}_{k}] $ and the $L$-Lipschitzness of $f_{i}$.
	Furthermore, we have $ \left\| \alpha_{k+1} G(\mathbf{x}_{k+1}) - \alpha_{k} G(\mathbf{x}_{k+1}) \right\|_{C}^{2} \leq L^{2} \max\{ \alpha_{k+1}^{2}, \alpha_{k}^{2} \} \left\| \mathbf{x}_{k+1} - \mathbf{x}_{k} \right\|_{C}^{2}$. 
	For $\|\mathbf{x}_{k+1} - \mathbf{x}_{k} \|_{C}^{2}$, we first derive
	\begin{align*}
		& \mathbf{x}_{k+1} - \mathbf{x}_{k} \\
		=& (\mathbf{R}_{\gamma} - I) (\mathbf{x}_{k} - \mathbf{1}\bar{\mathbf{x}}_{k}^{T}) - (\mathbf{v}_{k} - \pi_{\mathbf{C}}\hat{\mathbf{v}}_{k}^{T}) - \pi_{\mathbf{C}} \hat{\mathbf{v}}_{k}^{T} \\
		&+ \phi \mathbf{R} \boldsymbol{\zeta}_{k} - \gamma \mathbf{C} \boldsymbol{\xi}_{k},
	\end{align*}
	and then,
	\begin{align} \label{eq: v_3}
		& \mathbb{E}\left[\left\|\mathbf{x}_{k+1} - \mathbf{x}_{k} \right\|_{C}^{2} | \mathcal{F}_{k} \right] \nonumber \\
		\leq & 3 \left\| \mathbf{R}_{\gamma} - I \right\|_{C}^{2} \left\| \mathbf{x}_{k} - \mathbf{1}\bar{\mathbf{x}}_{k}^{T} \right\|_{C}^{2} + 3 \left\| \mathbf{v}_{k} - \pi_{\mathbf{C}}\hat{\mathbf{v}}_{k}^{T} \right\|_{\mathbf{C}}^{2} \nonumber \\
		&+ 3 \left\| \pi_{\mathbf{C}} \right\|_{C}^{2} \left\| \hat{\mathbf{v}}_{k}^{T} \right\|_{C}^{2}  +\phi^{2} \left\| \mathbf{R} \right\|_{C}^{2}Nm \theta_{\zeta,k}^{2} \nonumber \\
		&+ \gamma^{2}\left\| \mathbf{C} \right\|_{C}^{2} Nm \theta_{\xi,k}^{2} \nonumber
		\\
		\leq & 3 \sigma_{C,R}^{2}\left[ \left\| \mathbf{R}_{\gamma} - I \right\|_{C}^{2} + 2L^{2}N \alpha_{k}^{2}\left\| \pi_{\mathbf{C}} \right\|_{C}^{2} \right] \left\| \mathbf{x}_{k} - \mathbf{1}\bar{\mathbf{x}}_{k}^{T} \right\|_{R}^{2} \nonumber \\
		&+ 3 \left\| \mathbf{v}_{k} - \pi_{\mathbf{C}}\hat{\mathbf{v}}_{k}^{T} \right\|_{\mathbf{C}}^{2} + 6 \left\| \pi_{\mathbf{C}} \right\|_{C}^{2} \sigma_{C,R}^{2} \alpha_{k}^{2} \mathbb{E}[\left\| \nabla f(\bar{\mathbf{x}}_{k})\right\|_{2}^{2}] \nonumber \\
		& + \phi^{2} \left\| \mathbf{R} \right\|_{C}^{2}Nm \theta_{\zeta,k}^{2} + \gamma^{2}\left\| \mathbf{C} \right\|_{C}^{2} Nm \theta_{\xi,k}^{2}.
	\end{align}
	Combining~\eqref{eq: v_1}--\eqref{eq: v_3} , we have
	\begin{align}
		&\mathbb{E}[\left\| \mathbf{v}_{k+1} - \pi_{\mathbf{C}}\hat{\mathbf{v}}_{k+1}^{T} \right\|_{C}^{2}] \\
		\leq & \left( \frac{1+\sigma_{C}^{2}}{2} + \mathbf{a}_{5}\max\{ \alpha_{k+1}^{2}, \alpha_{k}^{2} \} \right) \left\| \mathbf{v}_{k+1} - \pi_{\mathbf{C}}\hat{\mathbf{v}}_{k+1}^{T} \right\|_{C}^{2} \nonumber \\
		&+ (\mathbf{a}_{3} + \mathbf{a}_{4} \alpha_{k}^{2}) \max\{ \alpha_{k+1}^{2}, \alpha_{k}^{2} \} \left\| \mathbf{x}_{k} - \mathbf{1}\bar{\mathbf{x}}_{k}^{T} \right\|_{R}^{2} \nonumber \\
		&+ B_{21} \alpha_{k}^{2} \max\{ \alpha_{k+1}^{2}, \alpha_{k}^{2} \} \mathbb{E}[\left\| \nabla f(\bar{\mathbf{x}}_{k})\right\|_{2}^{2}] \nonumber \\
		&+ B_{22} \max\{ \alpha_{k+1}^{2}, \alpha_{k}^{2} \} \phi^{2}  \theta_{\zeta,k}^{2} 
		+ B_{23} \gamma^{2} \theta_{\xi,k}^{2}.
	\end{align}

\subsection{Proof of Theorem~\ref{thm: primal_con}}
Due to the strong convexity of $F$, the Lagrangian $\mathcal{L}(\mathbf{w}, x)$ given in~\eqref{eq: lang} is also strongly convex. Specifically, we have:
		\begin{align*}
			\mathcal{L}(\mathbf{w}^{*}, \bar{\mathbf{x}}_{k}) \geq & \mathcal{L}(\mathbf{w}_{k}, \bar{\mathbf{x}}_{k}) + \nabla_{\mathbf{w}}\mathcal{L}(\mathbf{w}_{k}, \bar{\mathbf{x}}_{k})^{T}(\mathbf{w}^{*} - \mathbf{w}_{k}) \\ 
			& + \frac{\mu}{2} \| \mathbf{w}^{*} - \mathbf{w}_{k} \|^{2}.
		\end{align*}
		Under Assumption~\ref{assum: convexity_slater}, the strong duality holds and $f^{*} = -F^{*} = -\mathcal{L}(\mathbf{w}^{*}, x)$. Therefore, we obtain:
		\begin{align*}
			f(\bar{\mathbf{x}}_{k}) - f^{*} =& \mathcal{L}(\mathbf{w}^{*}, \bar{\mathbf{x}}_{k}) - \inf_{\mathbf{w} \in \mathcal{W}_{i}} \mathcal{L}(\mathbf{w}, \bar{\mathbf{x}}_{k} ) \\
			=& \mathcal{L}(\mathbf{w}^{*}, \bar{\mathbf{x}}_{k}) - \mathcal{L}(\mathbf{w}_{k}, \bar{\mathbf{x}}_{k}) \\
			\geq & \frac{\mu}{2} \| \mathbf{w}^{*} - \mathbf{w}_{k} \|^{2},
		\end{align*}
		where the inequality follows from the first-order necessary condition for a constrained minimization problem, i.e., $-\nabla_{\mathbf{w}}\mathcal{L}(\mathbf{w}_{k}, \bar{\mathbf{x}}_{k})^{T}(\mathbf{w}^{*} - \mathbf{w}_{k}) \leq 0$.
		By rearranging terms, we have
		\begin{equation} \label{eq: pb}
			\mathbb{E}[\| \mathbf{w}_{k}- \mathbf{w}^{*} \|^{2}] \leq \frac{2}{\mu} (\mathbb{E}[f(\bar{\mathbf{x}}_{k})] - f^{*}).
		\end{equation}
		Since $\mathbb{E}[f(\bar{\mathbf{x}}_{k})] - f^{*}$ in~\eqref{eq: pb} converges to a finite value almost surely according to Theorem~\ref{thm: convergence}, it follows that $\mathbb{E}[\|\mathbf{w}_{k} - \mathbf{w}^{*} \|^{2}]$ also converges to a finite value almost surely.
	
	\subsection{Proof of Corollary~\ref{cor: finite}} \label{app: cor}
	First, for the step size and noise given in Corollary~\ref{cor: finite}, we have $\sum_{k=0}^{\infty}\theta_{\xi,k}^{2} = \theta_{\xi,0}^{2}\sum_{k=0}^{\infty}q_{\xi}^{2k} = \frac{\theta_{\xi,0}^{2}}{1-q_{\xi}} < \infty$, $\sum_{k=0}^{\infty}\theta_{\zeta,k}^{2} = \theta_{\zeta,0}^{2}\sum_{k=0}^{\infty}q_{\zeta}^{2k} = \frac{\theta_{\zeta,0}^{2}}{1-q_{\zeta}} < \infty$, $\sum_{k=0}^{\infty} \alpha_{k} = \alpha_{0}\sum_{k=0}^{\infty}q^{k} = \frac{\alpha_{0}}{1-q} < \infty$, $\sum_{k=0}^{\infty} \frac{\theta_{\xi,k}^{2}}{\alpha_{k}} = \frac{\theta_{\xi,0}^{2}}{\alpha_{0}} \sum_{k=0}^{\infty} \left(\frac{q_{\xi}^{2}}{q} \right)^{k} = \frac{\theta_{\xi,0}^{2}q}{\alpha_{0}(q-q_{\xi}^{2})} < \infty$, and $\sum_{k=0}^{\infty} \frac{\theta_{\zeta,k}^{2}}{\alpha_{k}} = \frac{\theta_{\zeta,0}^{2}}{\alpha_{0}} \sum_{k=0}^{\infty} \left(\frac{q_{\zeta}^{2}}{q} \right)^{k} = \frac{\theta_{\zeta,0}^{2}q}{\alpha_{0}(q-q_{\zeta}^{2})} < \infty$, which satisfies sufficient conditions in Theorem~\ref{thm: primal_con}. Therefore, under DP-DGT, $\|\mathbf{w}_{k} - \mathbf{w}^{*}\|^{2}$ is stochastically bounded.
	
	Then, we consider two non-decreasing nonnegative sequence $\{ \varphi_{k}^{\prime} \}$ and $\{ \eta_{k}^{\prime} \}$, iteratively evolving as follows:
	\begin{equation} \label{eq: ub}
		\begin{aligned}
			\varphi_{k+1}^{\prime} &= \varphi_{k}^{\prime} + \frac{\alpha}{\mu} \eta_{k}^{\prime} + \frac{\alpha\delta}{\mu}, \\
			\eta_{k+1}^{\prime} &= 2 \varphi_{k}^{\prime} + \eta_{k}^{\prime} + \frac{\alpha}{\mu} \eta_{k}^{\prime} + \frac{\alpha\delta}{\mu}, 
		\end{aligned}
	\end{equation}	  
	with $\varphi_{0}^{\prime} = 0$ and $\eta_{0}^{\prime} = 0$. Then, we have $\varphi_{k} \leq \varphi_{k}^{\prime}$ and $\eta_{k} \leq \eta_{k}^{\prime}$ according to~\eqref{eq: DTD}. Based on~\eqref{eq: ub}, $\varphi_{k+1}^{\prime} = \frac{\alpha}{\mu} \sum_{t=0}^{k} \eta_{t}^{\prime} + \frac{\alpha \delta}{\mu} (k+1) \leq \frac{\alpha}{\mu}(\delta + \eta_{k}^{\prime})(k+1)$, and thus $\varphi_{k}^{\prime} \leq \frac{\alpha}{\mu}(\delta + \eta_{k}^{\prime})k$. Hence, $\eta_{k}^{\prime}$ is increasing with $k$, indicating that $D_{\eta}$ is increasing with $K$.
	
	Under the conditions listed in Corollary~\ref{cor: finite}, one has $\bar{\alpha} = \alpha_{0}$, $k_{0} = 0$, and $K = 0$. Therefore, we derive that $D_{\eta} = \frac{2\alpha_{0}\delta}{\gamma\phi\mu - \alpha_{0}}$. Additionally, we have $D_{\alpha, \xi} = \frac{\alpha_{0} q_{\xi}}{\theta_{\xi, 0}(q_{\xi} - q)}$ and $D_{\alpha, \zeta} = \frac{\alpha_{0} q_{\zeta}}{\theta_{\zeta, 0}(q_{\zeta} - q)}$. Therefore, we obtain that
	\begin{equation*}
		\epsilon = \frac{\alpha_{0}\delta(\gamma \phi \mu + \alpha_{0})}{\gamma \phi \mu (\gamma \phi \mu - \alpha_{0})}\left( \frac{q_{\xi}}{\theta_{\xi,0}(q_{\xi} - q)} + \frac{\phi q_{\zeta}}{\theta_{\zeta,0}(q_{\zeta} - q)}  \right).
	\end{equation*}
    Regarding the convergence error bound, from inequality (24), we obtain:
\begin{align*}
& \mathbb{E}[f(\bar{\mathbf{x}}_{k+1})] -f^{*} \\
\leq &  \mathbb{E}[f(\bar{\mathbf{x}}_{k})] - f^{*}  + r_{3,k} \\
\leq & \mathbb{E}[f(\bar{\mathbf{x}}_{k})] - f^{*}  + \frac{r_{3,k}}{\alpha_{k}} \\
\leq & f(\bar{x}_{0}) - f^{*} + \sum_{s=0}^{k} \frac{r_{3,s}}{\alpha_{s}} \\
\leq & O\left(f(\bar{x}_{0}) - f^{*} + \sum_{s=0}^{k} \frac{\theta_{\zeta,s}^{2}}{\alpha_{s}} + \sum_{s=0}^{k} \frac{\theta_{\xi,s}^{2}}{\alpha_{s}} +  \sum_{s=0}^{k} \frac{X_{s}}{\alpha_{s}} + \sum_{s=0}^{k} \frac{V_{s}}{\alpha_{s}} \right). \\
\end{align*}
If the step size and noise satisfy the conditions in Corollary 2, we get:
\begin{align*}
& \mathbb{E}[f(\bar{\mathbf{x}}_{k+1})] -f^{*} \\
\leq &  \mathbb{E}[f(\bar{\mathbf{x}}_{k})] - f^{*}  + r_{3,k} \\
\leq & O\left(f(\bar{x}_{0}) - f^{*} + \sum_{s=0}^{k} \frac{\theta_{\zeta,s}^{2}}{\alpha_{s}} + \sum_{s=0}^{k} \frac{\theta_{\xi,s}^{2}}{\alpha_{s}} +  \sum_{s=0}^{k} \frac{X_{s}}{\alpha_{s}} + \sum_{s=0}^{k} \frac{V_{s}}{\alpha_{s}} \right) \\
\leq & O \left( f(\bar{x}_{0}) - f^{*} + \frac{1}{(q - q_{\zeta}^{2})(q - q_{\xi}^{2})(q-q_{c})(q-q_{r})(1-q)}   \right).
\end{align*}
Then, the convergence of the primal variables satisfies:
\begin{align*}
    & \| \mathbf{w}^{*} - \mathbf{w}_{k} \|^{2} \\
    \leq & \frac{2}{\mu} (f(\bar{\mathbf{x}}_{k}) - f^{*}) \\
    \leq & O \left( 1 + \frac{1}{(q - q_{\zeta}^{2})(q - q_{\xi}^{2})(q-q_{c})(q-q_{r})(1-q)}   \right).
\end{align*}

\end{document}